\documentclass[11.75pt,a4paper]{article}

\usepackage[a4paper,top=2cm,bottom=2cm,left=2cm,right=2.5cm,marginparwidth=1.75cm]{geometry}
\usepackage[T1]{fontenc}
\usepackage{bm}
\usepackage{tikz}
\usetikzlibrary{calc}
\usepackage{amsmath}
\usepackage{graphicx}
\usepackage{afterpage}
\usepackage{booktabs}
\usepackage{multirow}
\usepackage[bookmarksopen,bookmarksnumbered,colorlinks,linkcolor=blue,citecolor=blue]{hyperref}
\usepackage{apacite} 
\usepackage{natbib}
\usepackage{hyperref}

\usepackage{rotating}
\usepackage{float}
\usepackage{caption}
\usepackage{subcaption}
\usepackage{array}
\newcounter{covidsubfig}

\newcounter{swisssubfig}

\newcounter{decsubfig}

\usepackage{pdflscape}
\usepackage{geometry}   

\newcommand{\cmpanel}[5]{%
  \begin{minipage}[t]{0.235\linewidth}\centering
    \refstepcounter{#1}%
    \includegraphics[width=\linewidth]{#2}%
    \captionsetup{justification=centering,font=footnotesize}
    \caption*{Figure \csname the#1\endcsname: #3 confusion matrices, #4 data.}
    \label{#5}%
  \end{minipage}%
}

\usepackage[affil-it]{authblk}

\makeatletter

\renewcommand{\section}{\@startsection{section}{1}{\z@}%
  {-3.5ex \@plus -1ex \@minus -.2ex} 
  {2.3ex \@plus .2ex} 
  {\large\rmfamily\bfseries}} 

\renewcommand{\subsection}{\@startsection{subsection}{2}{\z@}%
  {-3.25ex \@plus -1ex \@minus -.2ex} 
  {1.5ex \@plus .2ex} 
  {\rmfamily\bfseries}} 
\makeatother

\title{\bfseries \normalsize Using machine learning metrics to provide deeper insights into the performance of choice models}

\author[1]{Lorenzo Mu\~{n}oz*}
\author[1]{Stephane Hess}
\author[1]{Thomas O. Hancock}
\author[1]{Georges Sfeir}

\affil[1]{Choice Modelling Centre and Institute for Transport Studies, University of Leeds, UK. \vspace{-0.2cm}}
\affil[*]{Corresponding author}
\date{\vspace{-5ex}}

\begin{document}
\maketitle

\section*{Abstract}\small

Machine learning (ML) techniques are increasingly drawing interest in the choice modelling (CM) field. The focus has primarily been on comparing the performance of these contrasting approaches or on improving behavioural insights for ML techniques, rather than translating ideas from one field into the other. In the present paper, we specifically focus on knowledge transfer from ML into CM in the context of model performance evaluation. In CM, model performance is typically evaluated using log-likelihood and related indicators, which are aggregate fit metrics that focus on overall fit. Conversely, in ML, the focus is on alternative-level misclassifications and correct classifications, which provide a more nuanced view of the results. To bridge these approaches, we explore the use of a probabilistic version of the confusion matrix, which reports the average probability of the model predicting each alternative, conditional on which alternative was observed to be chosen, across all choice tasks. This enables the computation of probabilistic ML metrics for both classic choice models and ML algorithms. We analyse model performance jointly in terms of overall fit and alternative-level predictions. Our findings demonstrate that models with similar log-likelihood can exhibit substantially different confusion matrices, revealing different probability patterns that aggregate metrics cannot capture. This framework identifies where models systematically `confuse' alternatives, highlighting trade-offs between alternatives, and potentially guiding model specification. Furthermore, evaluating these matrices and metrics out-of-sample reveals alternative-level prediction shifts that significantly impact forecasting performance.

\textbf{Keywords}: Model performance, probabilistic machine learning metrics, confusion matrices. 

\section{Introduction}

Random Utility Maximisation (RUM) models \citep{mcfadden1981econometric, ortuzar} have been the dominant framework for analysing and forecasting choice behaviour for the past five decades \citep{hess2024handbook}. These models rely on economic theory to represent decision-making and have been widely applied within various fields, including transportation, health, marketing, and environmental economics. In recent years, the rapid expansion of digital technologies, the availability of large-scale behavioural data, and major advances in machine learning (ML) have motivated researchers to revisit and extend conventional choice modelling (CM) approaches and standard practices. This has led to interest in integrating the strengths of both econometric and ML traditions \citep{hillel2021systematic}, reflected in a growing body of work looking at the differences in modelling approaches, estimation strategies, and evaluation practices between ML and CM, and possible cross-fertilisation \citep{vancrane2022,HESS2026105846}.

Recent studies demonstrate several potential benefits of ML methods for CM applications, including the ability to analyse large datasets, incorporate different types of data (e.g., text and images), assist model specification, improve predictive performance, and capture more complex patterns in explanatory variables, including non-linearities and interactions \citep{sun2018,reslogit,wang2020b,ortelli2021,vancrane2022,vancrane2025}. 

Much of the current literature on comparing, contrasting, or combining CM and ML focuses on predictive performance, robustness, and interpretability, and many studies report that ML methods achieve superior predictive accuracy \citep{wang2024}. Notwithstanding the fact that focusing exclusively on predictive accuracy may be misleading or even detrimental in certain applications \citep{naser2025}, a key question arises as to how to evaluate predictive accuracy per se. 

In CM, model performance has historically relied on log-likelihood and related/derived indicators---including $\rho^2$, Akaike Information Criterion (AIC), and Bayesian Information Criterion (BIC). While theoretically well-founded, these criteria provide only a partial view of model performance, focusing on aggregate fit and correct predictions alone---with only the probability for the chosen alternative contributing to these metrics. Such metrics can mask systematic weaknesses, particularly across alternative-specific performance, or in imbalanced choice contexts, i.e., datasets where some alternatives are chosen substantially less often than others, because they fail to reveal how probability mass is distributed across non-chosen alternatives (a point we return to in \autoref{subsec:traditional_CM}).

In contrast, the ML literature has developed a rich set of evaluation metrics designed to assess predictive performance more comprehensively and from multiple perspectives. There is extensive focus on classification-based metrics, including confusion matrices (a common practice in ML to visualise and analyse results), and derived metrics---accuracy, sensitivity, specificity, and balanced accuracy. These metrics quantify both correct classifications and specific misclassification types, providing detailed insights into performance across alternatives. 

Of course, there are reasons for the differences in approaches. As mentioned above, in CM, the focus is typically on aggregate prediction. Conversely, in ML, the focus is mainly on individual-level classification accuracy. Thus, the choice of evaluation metrics is central in ML, determining how model performance is quantified and implicitly influencing the interpretation of feature importance. Indeed, there are several ways a model or its evaluation can go wrong, and a correct interpretation of the results allows the user to distinguish which instances to inspect \citep{lime}.

The extensive set of approaches used for model evaluation in ML provides an opportunity to deploy corresponding metrics in CM. This serves a dual purpose. Aside from providing further CM-specific insights into model performance, it will also facilitate more in-depth comparison of CM and ML results. However, before doing so, an additional distinction needs to be highlighted. While CM is theory-driven, ML offers an alternative data-driven approach to model choice behaviour, and its evaluation is strongly grounded in deterministic rather than probabilistic approaches. Part of the reason is that, in many ML applications, the aim is to use these predictions to obtain a yes/no output (e.g., passport gates, fraud detection, movie suggestions), where the highest-probability alternative is taken as the prediction and only its performance matters. This key difference helps explain why deterministic metrics have become dominant in ML, while in CM, these are seen as incomplete or even biased due to ignoring the probabilistic nature of the models \citep{train2009}. We further discuss this in \autoref{subsec:probabilistic_CM}. Before we proceed, a brief note on terminology is warranted. CM refers to alternatives, and the selection between them, while ML talks about classification, referring to methods as classifiers, and to alternatives or outcomes as classes. With the present paper being directed at choice modellers, we generally adopt the former terminology, noting that the methods themselves are readily transferable independent thereof.

This paper contributes to the existing literature by providing a framework that enables choice modellers to obtain a more nuanced understanding of model performance, whilst retaining probabilistic interpretations. Specifically, we introduce a probabilistic confusion matrix into the CM field, complementing traditional evaluation criteria and overcoming the limitations of deterministic metrics commonly used in ML. From this matrix, we derive complementary probabilistic performance metrics and compare them with conventional CM performance indicators (i.e., log-likelihood), assessing a range of CM and ML models. Extending these ideas from ML to CM offers the opportunity to better understand trade-offs between alternatives, i.e., false predictions from/towards competing alternatives \citep{yacouby2020}; account for probabilistic predictions for non-chosen alternatives; and ultimately guide improvements in both model structure and utility specification. By revealing where a model struggles, we provide a practical tool for targeting alternatives for which good prediction requires additional explanatory variables, interaction terms, or different functional forms. 

The remainder of the paper is structured as follows: \autoref{sec:metrics} reviews the traditional metrics for analysing performance in the CM field, \autoref{sec:methods} presents the typical ML approaches and develops a probabilistic extension of these ML metrics, \autoref{sec:case_studies} presents the case studies with the associated results, and \autoref{sec:conclusion} presents the conclusions and next steps. We also present the mathematical background for the estimated choice models and ML algorithms in \autoref{appendix:modelling_background}, and the overfitting check in \autoref{appendix:overfitting}.

\section{Traditional performance metrics in choice modelling}
\label{sec:metrics}

The CM literature uses a wide range of different model structures, including (but not limited to) models using the paradigm of RUM. Independent of the model used, let us assume that we have a sample of $N$ people\footnote{Here, $N$ relates to the sample actually used in estimation, i.e., excluding any hold-out dataset aside for validation.}, where person $n$ faces $T_n$ choice tasks, and where $J_{nt}$ alternatives are available to person $n$ in choice task $t$. We denote by $O=\sum_{n=1}^N T_n$ the total number of choice observations in the sample. Let us further define the chosen alternative for person $n$ in the choice task $t$ as $y_{nt}$, with $y_n=\left<y_{n1},\hdots,y_{nT_n}\right>$ giving the observed sequence of choices for person $n$. The analyst specifies a model that uses a vector of model parameters $\beta$ (of length $K$), with $\beta=\left<\beta_1,\hdots,\beta_K\right>$. The log-likelihood (LL) function for this model is then given by:
\begin{equation}\label{eq:LL}
LL\left(\beta\right)=\sum_{n=1}^N \ln L_{n}\left(y_{n}\mid\beta\right),
\end{equation}

where $L_{n}\left(y_{n}\mid\beta\right)$ is the likelihood of the observed sequence of choices for person $n$. The specific functional form for $L_{n}\left(y_{n}\mid\beta\right)$ will vary across models, but will be a function of the probability assigned by the model to the chosen alternative in each given choice situation, say $P_{n}\left(y_{nt}\mid\beta\right)$ for person $n$ in task $t$. The use of maximum likelihood estimation of a discrete choice model on a given sample yields the maximum likelihood estimator (MLE) $\hat{\beta}$ as:

\begin{equation}\label{eq:mle}
\hat{\beta}=\underset{\beta}{\arg\max} \,LL\left(\beta\right).	
\end{equation}
The LL at convergence, say $LL\left(\hat\beta\right)$, is a key input to model selection and model evaluation in CM. The most prominent example is the likelihood ratio (LR) test, which compares the fit of a general model and its constrained version, both estimated on the same data. Let $\hat{\beta}$ be the parameters obtained in the unconstrained estimation, while $\tilde{\beta}$ refers to the constrained case, e.g. where we constrain a single parameter, say $\beta_k=0$. If the null hypothesis $H_0$ of no difference between the models is true, the test statistic converges asymptotically to a $\chi^2$ distribution with $d$ degrees of freedom, where $d$ is the number of additional parameters in the general model. 

We then have:
\begin{equation}\label{eq:LR}
LR=2\left(LL_{\hat{\beta}}-LL_{\tilde{\beta}}\right)\sim \chi^2_d \quad \text{under } H_0.
\end{equation}
The LR test can only be used for nested comparisons, i.e., where one model is a more general version of another model. To allow for other comparisons, choice modellers also routinely report a number of additional measures, broadly described as goodness of fit criteria. 

The first example of these is the adjusted rho-squared ($\bar{\rho^2}$) measure, calculated as:

\begin{equation}
    \bar{\rho^2} = 1-\frac{LL(\hat{\beta})-K}{LL(0)},
\label{eq:rho2}
\end{equation}

where $LL(\hat{\beta})$ is again the LL at convergence, $LL(0)$ is the LL of a purely random model, and $K$ is the number of estimated parameters, penalising the model for additional complexity. This measure provides a conceptual interpretation of the proportion of variation explained by the model, and captures the relative improvement from a null model to a fitted model \citep{mokhtarian2016rho}. An alternative formulation of this metric compares $LL(\hat{\beta})$ to the fit of a constants-only model. In addition, $\bar{\rho^2}$ serves as the input to a formal test to compare two non-nested models, as put forward by \citet{benakivaswait}.

Another example based on LL is the Akaike Information Criterion (AIC) \citep{akaike1974}. While AIC is asymptotically efficient in non-parametric settings as sample size $N$ increases, it loses both efficiency and consistency in parametric scenarios where multiple candidate models are evaluated \citep{zhang2023}. It is calculated as:

\begin{equation}
    AIC = -2LL(\hat{\beta})+2K.
\end{equation}

Finally, the Bayesian Information Criterion (BIC) also includes the number of observations (not individuals) $O$ \citep{bayesianinfo}, thus penalising each additional parameter more with larger samples. It is consistent in a parametric scenario (in terms of the samples' probability distribution), but not in a non-parametric scenario \citep{zhang2023}. 

\begin{equation}
    BIC = -2LL(\hat{\beta})+K\ln(O).
\end{equation}

Both AIC and BIC require the number of estimated parameters $K$, which constitutes a fundamental difference for ML algorithms, as this concept lacks a direct equivalent in many ML architectures. Hence, prediction performance in ML algorithms typically relies on different metrics that do not account for $K$. Furthermore, $LL(\hat{\beta})$ and by implication AIC and BIC, cannot be compared across datasets as they depend on the number of alternatives as well as the sample size. The former also applies to $\bar{\rho^2}$, yet it can give some indication of the degree of randomness of a model, independent of sample size. 

Additional criteria for analysing model performance in the CM field rely on post-estimation calculations based on predicted probabilities for individual alternatives in individual choice situations, say $P_{int}\left(\hat{\beta}\right)$ for alternative $i$ in choice situation $t$ for person $n$. A common example includes comparing predicted market shares, i.e. $\widehat{MS}_i=\frac{1}{O}\sum^N_{n=1}\sum^{T_n}_{t=1}{P_{int}\left(\hat{\beta}\right)}$, and observed market shares, i.e. $MS_i=\frac{1}{O}\sum^N_{n=1}\sum^{T_n}_{n=1}y_{int}$, where $y_{int}=1$ if and only if alternative $i$ was chosen in that choice situation. However, this comparison alone does not provide a quantitative measure to evaluate the level of agreement between predictions and observations \citep{parady2021}, as evaluating aggregate market shares allows large prediction errors across observations to cancel out, therefore masking observation-level performance. 

A metric that is increasingly gaining popularity in CM is the Brier score \citep{krueger2021,wamhoff2024,lukawska2025}, defined as the mean square difference between predicted individual choice probabilities and observed choices, given by:

\begin{equation}
     Brier\ score=\frac{1}{O}\sum_{n=1}^N\sum_{t=1}^{T_n}\sum_{i=1}^{J_{nt}} (y_{int} - P_{int}\left(\hat{\beta}\right))^2.
\end{equation}

Brier scores are exclusively minimised by larger predicted choice probabilities for the alternatives observed to be chosen, providing a strictly proper scoring rule \citep{brier}, and take into account the predicted choice probabilities of the whole choice set. The lower the Brier score value, the better the model performs in terms of prediction.

Other performance metrics, often used in ML as well, are Sum of Squared Error (SSE), Mean Squared Error (MSE), Root Sum Squared Error (RSSE), Mean Absolute Error (MAE), and Mean Absolute Percentage Error (MAPE). Typically, these errors are considered as the differences between the observed choices $y_{int}$ and predicted probabilities $P_{int}(\hat\beta)$. These metrics differ in how they penalise errors: squared error metrics (e.g., SSE and MSE) place greater weight on large deviations, whereas absolute error metrics (e.g., MAE) treat all deviations proportionally. Percentage-based metrics (e.g., MAPE) express errors relative to the magnitude of the observed values. Still, when and how to use information criteria appropriately is a complex problem, and the uncertainty of model selection should be properly assessed \citep{zhang2023}. A summary of the aforementioned metrics is presented in \autoref{tab:metrics_summary}.

\begin{sidewaystable}
    \centering
    \normalsize
    \renewcommand{\arraystretch}{1.25}
    \caption{Summary of typical performance CM metrics with inputs.}
    \begin{tabular}{
        >{\centering\arraybackslash}p{3.0cm} |
        >{\centering\arraybackslash}p{5.0cm} |
        >{\centering\arraybackslash}p{5.0cm} |
        >{\centering\arraybackslash}p{7.0cm}}
    Name & Inputs & Equation & Key Benefit \\ \hline
    
    Log-likelihood
    & Choices ($y_{in}$) and parameters ($\beta$) 
    & {\small $LL\left(\beta\right)=\sum_{n=1}^N \ln L_{n}\left(y_{n}\mid\beta\right)$}
    & Core CM estimation criterion; aggregate measure based on correct predictions. \\ [8pt]
    
    Market share recovery
    & Predicted probabilities (${P}_{int}(\widehat\beta)$) 
    & {\small $\widehat{MS}_i=\frac{1}{O}\sum^N_{n=1}\sum^{T_n}_{t=1}{P_{int}\left(\hat{\beta}\right)}$}
    & Useful for forecasting and policy evaluation. \\ [8pt]
    
    $\bar{\rho^2}$ 
    & $LL(\hat\beta)$, $LL(0)$, and number of parameters ($K$) 
    & {\small     $\bar{\rho^2} = 1-\frac{LL(\hat{\beta})-K}{LL(0)}$}
    & Common for model comparison. \\ [8pt]

    LR 
    & Log-likelihood of an unconstrained model ($LL_{\hat{\beta}}$) and of a constrained model ($LL_{\tilde{\beta}}$) 
    & $LR=2\left(LL_{\hat{\beta}}-LL_{\tilde{\beta}}\right)$ 
    & Useful to compare models with different levels of complexity.  \\ 
    
    AIC 
    & $LL(\hat{\beta})$ and number of parameters ($K$) 
    & {\small $AIC = -2LL(\hat{\beta}) + 2K$}
    & Derived from log-likelihood, while penalising complexity. \\ [8pt]
    
    BIC 
    & $LL(\hat{\beta})$, number of parameters ($K$), and of observations ($O$) 
    & {\small $BIC = -2LL(\hat{\beta}) + K\ln(O)$}
    & Stronger penalty for complexity, especially for large datasets.\\[8pt]
    
    Brier Score 
    & Choices ($y_{int}$) and predicted probabilities ($P_{int}$) 
    & {\tiny$\frac{1}{O}\sum_{n=1}^N\sum_{t=1}^{T_n}\sum_{i=1}^{J_{nt}} (y_{int} - P_{int}(\hat\beta))^2$}
    & Complementary metric accounting for full probability vectors. \\ [8pt]
    
    SSE 
    & Choices ($y_{int}$) and predicted probabilities ($P_{int}(\hat\beta)$) 
    & {\small $\sum_{n=1}^N\sum_{t}^{T_n}\sum_{i=1}^{J_{nt}} (y_{int} - P_{int}(\hat\beta))^2$}
    & Measures total (unscaled) prediction error, heavily penalising large mispredictions due to its quadratic structure. \\ [8pt]

    MSE 
    & Choices ($y_{int}$) and predicted probabilities ($P_{int}(\hat\beta)$)  
    & {\small $\frac{1}{O}\sum_{n=1}^N\sum_{t}^{T_n}\sum_{i=1}^{J_{nt}} (y_{int} - P_{int}(\hat\beta))^2$}
    & Normalises squared predictions by sample size. \\ [8pt]

    RSSE 
    & Choices ($y_{int}$) and predicted probabilities ($P_{int}(\hat\beta)$)  
    & {\tiny $\sqrt{\sum_{n=1}^N\sum_{t}^{T_n}\sum_{i=1}^{J_{nt}} (y_{int} - P_{int}(\hat\beta))^2}$}
    & Squared root of SSE, easier to interpret.\\ [8pt]

    MAE 
    & Choices ($y_{int}$) and predicted probabilities ($P_{int}(\hat\beta)$)  
    & {\small $\frac{1}{O}\sum_{n=1}^N\sum_{t}^{T_n}\sum_{i=1}^{J_{nt}} |y_{int} - P_{int}(\hat\beta)|$}
    & Less sensitive to outliers than MSE, given its linear structure. \\ [8pt]

    MAPE 
    & Observed ($MS_{i}$) and predicted market shares ($\hat{MS}_{i}$) 
    & {\small $\frac{1}{J} \sum_{i=1}^J \left| \frac{MS_i - \widehat{MS}_i}{MS_i} \right|$}
    & Easy to interpret, useful for forecasting. \\

    \label{tab:metrics_summary}
    \end{tabular}
\end{sidewaystable}

In general, more parsimonious models are preferred over less parsimonious ones, as they fit the data adequately and capture the essential behavioural mechanisms \citep{akaike1974, bayesianinfo, hensher2015}. Parsimony ensures that a model explains the observed behaviour using the smallest number of parameters necessary, reducing the risk of overfitting and improving the model's ability to generalise to new data \citep{posada2004}. Simpler models are also easier to interpret and avoid issues of parameter instability. Hence, if two models perform equally in terms of the statistical evaluation criteria, the one that includes fewer variables is more likely to be chosen. Behavioural interpretability also plays a central role, as models are often used for forecasting and economic appraisal \citep{fox2014}. Thus, parameter magnitudes, signs, and behavioural plausibility are also considered. Furthermore, marginal rates of substitution, such as the value of time or other willingness-to-pay measures, and other economic outcomes, such as elasticities, are also frequently analysed. Model criteria, such as correct description and interpretation of the observed behaviour, reproduction of observations, and generalisation to choice behaviour in the same scenario, are also key to a proper specification \citep{deluca2016}.

\section{Confusion matrices and machine learning metrics} 
\label{sec:methods}

In this section, we first summarise the standard evaluation criteria used in ML. We introduce the binary confusion matrix and its components in \autoref{subsec:traditional_CM}, and then extend the construction to a multi-alternative setting in \autoref{subsec:multiclass_CM}. Finally, we present a probabilistic extension that replaces deterministic predictions with full probability vectors in \autoref{subsec:probabilistic_CM}, enabling the computation of probabilistic ML metrics---accuracy, sensitivity, specificity, and balanced accuracy.

\subsection{Traditional confusion matrices}
\label{subsec:traditional_CM}

The analogue of CM in ML is supervised classification, where the model is trained to predict the class of a given input. As mentioned earlier, in this analogy, the `class' in an ML context corresponds to the alternative $i$ in CM, while the `features' correspond to the attributes $x_{int}$. Many, though not all, ML models produce probabilistic outputs and are trained with probabilistic loss functions (e.g., cross-entropy), but model performance is often reported using deterministic metrics---computed by assigning each observation to the alternative with the highest predicted probability \citep{azamali2023}. 

When evaluating model predictions, a standard diagnostic tool in the ML community is the confusion matrix, which originates from binary classification between two outcomes (alternatives), namely positive and negative. This specific example is common in health applications, as many case studies seek to detect whether a condition is present (positive) or absent (negative). Here, the columns are the predicted alternatives and rows are the observed alternatives. The diagonal elements report the counts of correctly classified observations for each alternative, whereas the off-diagonal elements quantify misclassifications.

There are in that case four possible mutually exclusive outcomes, as shown in \autoref{fig:conf}, summarising the frequencies of correct and incorrect classifications. The first row corresponds to the cases where a positive outcome is observed, with the opposite applying for the second row. We then have:

\begin{itemize}
    \item True Positive (TP): the model correctly predicts the `positive' alternative. 
    \item False Negative (FN): the model incorrectly predicts the `negative' alternative when the observed alternative is `positive'. 
    \item False Positive (FP): the model incorrectly predicts the `positive' alternative when the observed alternative is `negative'. 
    \item True Negative (TN): the model correctly predicts the `negative' alternative.
\end{itemize}

\begin{figure}[htbp]
\centering
\begin{tikzpicture}[
    cell/.style={
        rectangle,
        draw=black,
        minimum width=3cm,
        minimum height=1.5cm,
        align=center,
        font=\small
    },
    label/.style={
        font=\small\bfseries,
        align=center
    }
]

\node[cell] (TP) at (0,0)
{$TP$};

\node[cell] (FN) at (3,0)
{$FN$};

\node[cell] (FP) at (0,-1.5)
{$FP$};

\node[cell] (TN) at (3,-1.5)
{$TN$};

\node[label] at (0,1.2) {Predicted\\Positive};
\node[label] at (3,1.2) {Predicted\\Negative};

\node[label, anchor=east] at (-1.6,0) {Observed\\Positive};
\node[label, anchor=east] at (-1.6,-1.5) {Observed\\Negative};

\node[label] at (1.5,2.0) {Predicted class};
\node[label, rotate=90] at (-4.0,-0.75) {Observed class};

\end{tikzpicture}
\caption{Binary confusion matrix.}
\label{fig:conf}
\end{figure}

By using the components of the confusion matrix (TP, FN, FP, and TN), a number of key metrics are commonly computed in ML. Some of the most widely used metrics in the ML field are accuracy, sensitivity, specificity, and balanced accuracy, which we introduce below. Using these evaluation criteria jointly gives a more detailed picture of model performance \citep{kononenko1991information}. Hence, a joint evaluation using multiple complementary metrics can be desirable.

Accuracy is the most commonly reported performance metric in ML. It measures the proportion of correctly classified cases relative to the total number of observations, and therefore provides a general summary of classification performance. The similarity with likelihood-based metrics is clear, as these also focus on the prediction of the actual observed outcomes.

\begin{equation}
Accuracy = \frac{TP+TN}{TP+TN+FP+FN}
\label{eq:accuracy}
\end{equation}

Many ML algorithms implicitly optimise accuracy by attempting to maximise the number of correctly classified observations, regardless of the alternative to which they belong. Because models naturally struggle to predict rarely chosen alternatives, these may yield few correct predictions, causing accuracy to be biased toward majority classes \citep{batista2004,sun2009}.
 
Sensitivity, also known as recall, measures the proportion of observed positives out of all the predicted positive cases. A high sensitivity indicates that the model is effective at recognising the positive alternative when it is observed. Unlike accuracy, sensitivity is not influenced by the number of TN and therefore provides a clearer assessment of how well a model detects the positive observations. 

\begin{equation}
Sensitivity = \frac{TP}{TP+FN}
\label{eq:sensitivity}
\end{equation}

However, prioritising sensitivity may come at the cost of increasing FP, thereby reducing precision \citep{terven2023}. Precision measures the proportion of correctly identified observations among all observations predicted as a given alternative, i.e., it is similar to sensitivity, but uses FP instead of FN in the denominator:

\begin{equation}
Precision = \frac{TP}{TP+FP}
\label{eq:precision}
\end{equation}

Specificity measures the proportion of observations that do not belong to a given alternative that are correctly identified as such. This metric is particularly important when FP are costly. For example, in medical diagnostics, incorrectly identifying a healthy individual as having a disease may trigger unnecessary tests, treatment, monitoring, and additional healthcare costs \citep{buckell2025}. 

\begin{equation}
Specificity = \frac{TN}{TN+FP}
\label{eq:specificity}
\end{equation}

Together, sensitivity and specificity provide complementary insights into model performance. Thus, while sensitivity focuses on minimising FN and ensuring that positive cases are detected, specificity focuses on minimising FP and avoiding the incorrect classification of negative cases. 

Balanced accuracy is defined as the arithmetic mean between sensitivity and specificity\footnote{Sensitivity and precision are sometimes combined into their harmonic mean, the $F_1$ score, and \cite{yacouby2020} propose a probabilistic extension of both. We focus instead on sensitivity, specificity, and their mean, as these are the quantities most directly connected to the recovery of market shares that is central to CM.}. It combines both positive and negative classifications into a single measure, reflecting a model's ability to correctly identify an alternative and distinguish it from competing alternatives.

\begin{equation}
Balanced\ accuracy = \frac{Sensitivity+Specificity}{2}
\label{eq:balance}
\end{equation}

Choosing the appropriate performance metric is essential for success in ML models \citep{terven2023}, as different applications place different values on specific types of prediction errors. For example, in banking security software, it may be preferable to prioritise sensitivity, as failing to identify a fraudulent transaction (FN) can expose users and institutions to substantial financial losses. In such contexts, accepting a higher number of false alarms (FP) may be considered an acceptable trade-off. Similarly, in medical diagnostics, sensitivity is often prioritised because failing to detect a disease (FN) may delay treatment and lead to severe consequences for the patient. While FP are also undesirable, they typically result in additional tests or examinations rather than missed care. Hence, the relative importance of sensitivity and specificity therefore depends on the consequences associated with FP and FN in the application of interest.

A natural question is why we consider multiple metrics when accuracy is commonly reported as the primary performance metric. The answer is that these metrics are complementary, rather than mutually exclusive. A particular reason for the interest in using multiple metrics is that of imbalance in datasets, i.e. where specific alternatives are chosen much more frequently than others. Consider a dataset in which 95\% of observations belong to a single alternative: achieving 95\% accuracy is guaranteed if the model simply predicts that alternative for everyone. 

Imbalanced datasets, i.e., where some alternatives are severely underrepresented, pose a particular challenge for ML algorithms because it is difficult to outperform such a baseline when the majority alternative dominates \citep{forman2003}. Consequently, accuracy is not sufficient when the dataset is imbalanced \citep{chawla2005}. In deterministic settings, accuracy systematically favours overrepresented classes because each observation is assigned exclusively to the alternative with the highest predicted probability, yielding misleading conclusions about model performance \citep{buda2018}. While probabilistic evaluation partially alleviates this issue by accounting for the full distribution of predicted probabilities, alternative imbalance remains challenging whenever model performance is assessed through discrete classifications. In such cases, a model can achieve high accuracy by correctly predicting the majority alternatives while performing poorly for low-share alternatives. When a model sees only a handful of examples for a minority alternative, it struggles to learn its underlying structure and instead gravitates toward the majority alternatives \citep{tornetta2021}.

Moreover, sensitivity is particularly useful for evaluating classification performance in imbalanced datasets \citep{hagenauer2017}, as it is not affected by the relative frequency of the alternative. Consequently, a model may achieve high overall accuracy while exhibiting low sensitivity for minority alternatives, revealing weaknesses that accuracy can conceal. Balanced accuracy addresses this limitation by combining sensitivity and specificity into a single metric. In doing so, it assigns equal importance to correctly identifying positive and negative cases, regardless of choice frequency. As a result, balanced accuracy provides a more informative assessment of performance when the number of positive and negative observations differs substantially across alternatives \citep{kim2021}. More generally, considering multiple metrics is particularly important in imbalanced datasets, where individual metrics may provide misleading signals \citep{yacouby2020}.

Another limitation of relying on a single metric is the possibility of metric invariance \citep{sokolova2009}, where a metric's value remains unchanged despite changes in the confusion matrix. For instance, sensitivity is invariant to changes in TN, while specificity is invariant to changes in TP. Consequently, relying on either metric on its own could fail to reflect shifts in predictive performance. Therefore, in our results, we report accuracy, sensitivity, specificity, and balanced accuracy all at the same time.

\subsection{Multi-alternative confusion matrices}
\label{subsec:multiclass_CM}

In the binary case, the confusion matrix decomposes predictions into four mutually exclusive cases, where our illustration has focused on positive and negative outcomes. The confusion matrix can be extended to a multi-alternative setting simply by moving away from a $2x2$ matrix, having one row (and one column) per alternative, treating each alternative $i$ as `positive', and aggregating the remaining competing alternatives as `negative'. In the multi-alternative confusion matrix, TP for alternative $i$ are in the diagonal entry $(i,i)$; FN for $i$ are in the off-diagonal entries of row $i$ (i.e., observations for $i$ predicted as another alternative); FP for $i$ are the off-diagonal elements of column $i$ (i.e., other alternatives predicted as $i$); and TN for $i$ comprise all cells outside row $i$ and column $i$. This approach preserves the possibility of binary definitions for each alternative, by summing the off-diagonals in a row to get FN, and summing the off-diagonals in a column to get FP, but at the same time enables alternative-specific analysis. 

To illustrate this, imagine a case with three alternatives, car (C), bus (B) and walk (W). The first diagonal element would be the true predictions for car, let's say $TC$, the second would be the true predictions for bus, $TB$, and the third one the true predictions for walk, $TW$. The off-diagonal elements correspond to misclassifications. For example, if car is observed to be chosen, but bus is predicted, we have $FB\mid C$. We show an example of this in \autoref{fig:conf_multi}.

\begin{figure}[htbp]
\centering
\begin{tikzpicture}[
    cell/.style={
        rectangle,
        draw=black,
        minimum width=2.4cm,
        minimum height=1.2cm,
        align=center,
        font=\small
    },
    label/.style={
        font=\small\bfseries,
        align=center
    },
    note/.style={
        font=\footnotesize,
        align=center
    }
]

\node[cell] (CC) at (0,0) {$TC$\\[-1mm]\footnotesize true car};
\node[cell] (BC) at (2.4,0) {$FB\mid C$\\[-1mm]\footnotesize bus predicted\\car observed};
\node[cell] (WC) at (4.8,0) {$FW\mid C$\\[-1mm]\footnotesize walk predicted\\car observed};

\node[cell] (CB) at (0,-1.2) {$FC\mid B$\\[-1mm]\footnotesize car predicted\\bus observed};
\node[cell] (BB) at (2.4,-1.2) {$TB$\\[-1mm]\footnotesize true bus};
\node[cell] (WB) at (4.8,-1.2) {$FW\mid B$\\[-1mm]\footnotesize walk predicted\\bus observed};

\node[cell] (CW) at (0,-2.4) {$FC\mid W$\\[-1mm]\footnotesize car predicted\\walk observed};
\node[cell] (BW) at (2.4,-2.4) {$FB\mid W$\\[-1mm]\footnotesize bus predicted\\walk observed};
\node[cell] (WW) at (4.8,-2.4) {$TW$\\[-1mm]\footnotesize true walk};

\node[label] at (0,1.0) {Predicted\\Car};
\node[label] at (2.4,1.0) {Predicted\\Bus};
\node[label] at (4.8,1.0) {Predicted\\Walk};

\node[label, anchor=east] at (-1.5,0) {Observed\\Car};
\node[label, anchor=east] at (-1.5,-1.2) {Observed\\Bus};
\node[label, anchor=east] at (-1.5,-2.4) {Observed\\Walk};

\node[label] at (2.4,1.75) {Predicted alternative};
\node[label, rotate=90] at (-4.0,-1.2) {Observed alternative};

\end{tikzpicture}
\caption{Multi-alternative confusion matrix for three alternatives: car, bus, and walk. Diagonal elements are correct classifications; off-diagonal elements are misclassifications. For example, $FB\mid C$ denotes a bus prediction when car was observed.}
\label{fig:conf_multi}
\end{figure}

The confusion matrix has several advantages when applied to CM. First, it provides a clear visual summary of the agreement between observed choices and model predictions, showing how predicted elements are distributed across alternatives. Second, it highlights alternatives for which the model systematically struggles, as the off-diagonal elements quantify the extent to which the model confuses one alternative for another. For example, in a multi-alternative setting, frequent misclassifications of an alternative as another alternative indicate either strong similarity between these choices or insufficient discrimination in model specification. Third, the confusion matrix provides direct insights into misclassification patterns, where persistent tendencies can reveal utility misspecification, such as omitted variables, poorly scaled attributes, or inappropriate functional forms. For instance, if bus observations receive a high predicted probability for rail but not for walk, the model is implicitly treating rail as the closest substitute, providing evidence that can inform the specification of a more flexible CM. Finally, it can help diagnose violations of structural assumptions, such as the independence of irrelevant alternatives: large probability misallocation between specific pairs of alternatives may indicate that the misclassification patterns implied by a specific model are inconsistent with those observed in the data. These insights are examined in detail across three datasets (see \autoref{sec:case_studies}).

\subsection{Probabilistic confusion matrices}
\label{subsec:probabilistic_CM}

Traditional confusion matrices in the ML field are typically computed using discrete predictions rather than the full vector of predicted probabilities. In CM, however, such `hit-rates' are frequently criticised as they implicitly assume that the decision-maker deterministically selects the alternative with the highest probability, thereby neglecting the model's probabilistic nature. Moreover, relying solely on deterministic metrics may fail to capture important aspects of multi-alternative decision problems, particularly when probability calibration and behavioural interpretation are central.

A similar argument was presented by \cite{train2009}: in a binary model, if an individual has a 99\% probability of choosing $i$ and a 1\% probability of choosing $j$, deterministic metrics will treat the outcomes the same as an individual that has a 51\% probability of choosing $i$ and a 49\% probability of choosing $j$, ignoring all nuances in the probabilities. The implications are worse still for multi-alternative models, where all the probabilities of the non-chosen alternatives would practically be ignored. For instance, in a three-alternative case, if an alternative is classified into its true alternative with a probability of 50\%, a deterministic approach ignores how the remaining 50\% is distributed. This could be concentrated in a single competing option (e.g., if for the remaining alternatives we had 49\% for one alternative and 1\% for the other), or evenly distributed (e.g., 25\% for each competing alternative). Indeed, analysing the probability distributions presents information on where the model performs well or poorly \citep{hillel2021systematic}. 

This issue is particularly relevant in CM, where datasets are often imbalanced, especially in revealed preference data. Assigning each observation to the alternative with the highest probability tends to under-represent low-probability alternatives and over-represent high-probability alternatives in the predicted outcomes. As a result, the predicted alternative shares may deviate substantially from the observed market shares, which is problematic in CM, where accurately reproducing aggregate shares is often a key objective \citep{hillel2021systematic}. In the worst case, the use of deterministic metrics is likely to yield zero predictions for minority classes.

Focussing on the issue from a model fit and model selection perspective, rather than prediction per se, the use of deterministic outputs implies using discontinuous metrics. This may result in very little difference in model performance metrics across models. Indeed, imagine a situation where there are many cases where the probability for a given alternative is already above 50\% in the base model. Any increases from 50\%, which would have no impact on which alternative is predicted to be chosen, lead to similar/equal performance in the subsequent metrics. This consequently results in a score that is neither differentiable nor strictly convex. This could entail uninformative metrics for model fitting when a continuous gradient is required, and even an inability to actually estimate the model, for instance when optimising an MNL model for deterministic accuracy. 

Some studies in the ML field have reported probabilistic outcomes for model evaluation. For example, \cite{cantarella2005} defined different cases, using percentages of `right' (individuals whose observed choices are given the maximum probability), `clearly right' (probability above 90\% to the chosen alternative), `clearly wrong' (probability over 90\% to an unchosen alternative), and `unclear' (cases where the model does not give a probability greater than 90\% to any alternative) for a multilayer feed-forward network. \cite{azamali2023} used probabilistic mean absolute percentage error (MAPE) of market shares in a vehicle ownership context. \cite{wang2013} developed a probabilistic confusion matrix for cross-entropy loss. \cite{tornetta2021} introduced an entropy-based confusion matrix for evaluating probabilistic classifiers. \cite{yacouby2020} proposed the use of probabilistic labels to extend standard performance metrics. Finally, \cite{markoulidakis2024} compared the probabilistic and deterministic confusion matrices. 

Notwithstanding these developments, the use of the probabilistic confusion matrix, along with probabilistic traditional ML metrics---such as accuracy, sensitivity, specificity, and balanced accuracy---remains largely unexplored in CM. This paper thus provides a probabilistic extension of the confusion matrix and associated ML metrics. Instead of assigning each observation to a single predicted class, the predicted choice probabilities are distributed across the corresponding row of the observed alternative. In this way, each observation contributes continuously to multiple cells according to the predicted probability vector. When aggregated across individuals, these probabilistic contributions replace the traditional deterministic counts and produce a matrix that captures the full shape of the model’s predicted probability distribution over the choice set. Formally, the element $C_{ij}$ in row $i$ and column $j$ of the probabilistic confusion matrix is defined as:

\begin{equation}
    C_{ij} = \sum_{n=1}^{N}\sum_{t=1}^{T_n} y_{int} P_{jnt},
    \label{eq:components}
\end{equation}

where $y_{int}$ is the binary indicator that equals $1$ if individual $n$ chose alternative $i$ in a choice task $t$ and 0 otherwise, and $P_{jnt}$ is the model's predicted probability vector for each alternative $j$ in the same task. Each entry $C_{ij}$ is thus a sum of predicted probabilities, i.e. a continuous, count-like quantity rather than a probability, equivalent to the predicted demand for given alternatives in cases where a given choice was observed. Availabilities are directly accommodated through zero probabilities being predicted by the model for unavailable alternatives.

The row sum $\sum_j C_{ij}$ recovers the observed count of alternative $i$, since the predicted probabilities sum to one within each observation. For interpretation and display, we report the row-normalised matrix:
 
\begin{equation}
    \widetilde{C}_{ij} = \frac{C_{ij}}{\sum_{j} C_{ij}} = \frac{C_{ij}}{MS_i},
    \label{eq:rownorm}
\end{equation}
 
whose entries are mean predicted probabilities conditional on the observed alternative: the diagonal $\widetilde{C}_{ii}$ is the mean probability assigned to the correct alternative when $i$ is chosen, and the off-diagonal $\widetilde{C}_{ij}$ is the mean probability mass the model places on $j$ when $i$ was chosen, which is the quantity that carries the substitution-pattern interpretation used throughout \autoref{sec:case_studies}. All confusion matrices in the case studies report $\widetilde{C}$.

Consequently, the components of the confusion matrix (TP, FP, TN, and FN) also become continuous, rather than discrete counts, so that the metrics of \autoref{eq:accuracy}--\autoref{eq:balance} are evaluated on these probabilistic cells. As a result, the diagonal entries accumulate probabilistic TP\footnote{A reader will note that, in the case of models without mixing, we recover the log‑likelihood by summing the logarithm of each per-observation diagonal contribution, i.e. $LL=\sum_{i=1}^J\sum_{n=1}^N\sum_{t=1}^T \ln (\widetilde{C}_{ii})$.}.

An advantage of the probabilistic confusion matrix is that it can distinguish between models with similar aggregate performance measures. Two models may achieve nearly identical values of log-likelihood, accuracy, or other summary metrics while allocating probabilities very differently across alternatives. Consequently, such models may exhibit markedly different confusion matrices, revealing different patterns of correct classifications and misclassifications. These differences are behaviourally meaningful, as they reflect distinct misclassification patterns between alternatives and may ultimately affect economic indicators derived from choice models, such as marginal rates of substitution and/or (cross-)elasticities. 

A further advantage of this approach arises in overfitting detection. When comparing training and testing confusion matrices we can analyse alternative-specific overfitting. While aggregate measures can indicate that predictive performance decreases out-of-sample, they do not reveal which alternatives drive this decrease. By contrast, confusion matrices show how predictive performance changes for each alternative individually. For example, a large reduction in the diagonal element of a specific alternative between the training and testing samples indicates that the model has learned patterns associated with that alternative that do not generalise well. Similarly, increases in particular off-diagonal elements reveal the alternatives towards which probability mass is reallocated out-of-sample.

We also expect to find an important structural property when evaluating alternative-specific performance. Let $D_i$ be the observed count, and $\widehat{D}_i$ be the predicted count for alternative $i$. The confusion matrix elements for alternative $i$ satisfy $D_i = TP_i + FN_i$ and $\widehat{D}_i = TP_i + FP_i$, hence:

\begin{equation}
    D_i - \widehat{D}_i = FN_i - FP_i.
    \label{eq:fp_fn}
\end{equation}

When a model correctly recovers market shares ($\widehat{D}_i = D_i$), such as with a Multinomial Logit model with a full set of alternative-specific constants, \autoref{eq:fp_fn} guarantees that $FP_i = FN_i$. Conversely, many ML algorithms do not enforce this, leading to market share discrepancies ($\widehat{D}_i \neq D_i$) and consequently $FP_i \neq FN_i$. Of course, when using deterministic predictions, this property no longer holds.

\section{Case studies}
\label{sec:case_studies}

In this section, we analyse three complementary case studies. The first is a stated preference (SP) survey on Covid vaccination choices. The second is a widely used SP dataset collected on long-distance mode choice in Switzerland, SwissMetro, with a three-alternative choice setting. Finally, we study mode choice using a revealed preferences (RP) dataset from the Decisions project, capturing daily travel choices across six alternatives in West Yorkshire, England. We estimate a total of eight models to effectively compare predictive performance between common CM and ML algorithms. 

Specifically, the CM specifications include a Multinomial Logit (MNL), a Nested Logit (NL), and models that were developed to better account for asymmetries in the probability distribution: Asymmetric Logit, Multinomial Scobit and Uneven Logit models \citep{brathwaite2018}. These models are detailed in \autoref{subsec_ap:choicemodels}. Based on the reviews of \cite{vancrane2022} and \cite{hillel2021systematic}, we select three of the most widely used ML classification algorithms: eXtreme Gradient Boosting (XGB), Artificial Neural Networks (ANN), and Random Forests (RF). For the estimation of all models, we used 80\% of the sample for training and 20\% for testing. For ML algorithms, we used a 20\% validation subset within the training set. We also ensured that the samples' market shares hold for both sub-samples and that we split for individuals, not observations, given the panel nature of the datasets. Hence, we prevent the same individual's choices from appearing in both the training and testing data to avoid bias \citep{hillel2020}. A hyperparameter search was conducted for all models to optimise their performance, and the details are described in \autoref{subsec_ap:ml}. No random parameter models were included---the advantages of these models in prediction are often small \citep{hess2025flexibility}, and there is a lack of an obvious counterpart model in ML.

\subsection{Covid vaccine dataset}
\label{subsec_case_covid}

This dataset comes from an SP survey conducted across all six inhabited continents to study vaccination preferences during the Covid-19 pandemic \citep{hess2022covid}. For our analysis, we use the subsample collected in the United Kingdom. This dataset is partly labelled in the sense that there is a choice between vaccination or no vaccination, and a choice between paid and free vaccines. Within a category (e.g. paid), the choice between vaccine A and B is unlabelled. The dataset is relatively imbalanced: 33.7\% of the choices correspond to free vaccine B, 29.6\% to paid vaccine B, 15.9\% to paid vaccine A, 13.9\% to free vaccine A, and 6.8\% to no vaccine. Attributes for this dataset include risk of infection, risk of serious illness, estimated protection duration, risk of mild side effects, risk of severe side effects, population coverage, exemption from international travel restrictions, waiting time for free vaccination, and a fee for paid priority access (see Figure 1 in \cite{hess2022covid}). 

We estimated the full set of CM and ML algorithms previously mentioned. For readability, the confusion matrices are row-normalised, such that the entries in each row sum to one. Consequently, each cell represents the mean probability assigned to a predicted alternative, conditional on the observed alternative, rather than the aggregate sum of probabilities across observations. We report overall fit using log-likelihood (LL), and provide alternative-specific metrics using accuracy, sensitivity, specificity, and balanced accuracy in \autoref{tab:metrics_covid}. We also present the confusion matrices for this dataset in \autoref{fig:covid_2A} through \autoref{fig:covid_2H}.

When comparing overall model fit, we observe substantial improvements in training log-likelihood when moving from the basic MNL model to the best-fitting model, XGB (from -13,322.35 to -8,955.02, a 33\% improvement). However, these gains do not translate into large improvements in testing log-likelihood (from -3,468.08 to -3,364.17, a 3\% improvement). Moreover, a first key finding from the confusion matrices is that, across all models and both data partitions, predictive performance is relatively weak in terms of correct classification. Indeed, diagonal elements remain quite low for all alternatives, indicating that these models struggle to differentiate between alternatives. Instead, substantial probabilities are assigned across competing alternatives. This is also highlighted by the low scores for overall accuracy, showing how these models struggle in imbalanced settings. 

An interesting insight arises when looking at specific alternatives, which have much higher scores than overall accuracy. These differences are driven by the structure of multi-alternative confusion matrices: while overall accuracy counts only correct predictions, alternative-specific accuracy incorporates TN across all unchosen alternatives. For example, free vaccine A observations are frequently assigned higher probabilities for the corresponding paid alternative, and similar misclassification patterns arise between free and paid vaccine B. Furthermore, the vast majority of no vaccine observations are misclassified as a vaccine, yielding a high number of FN. We also note how no vaccine has the highest alternative-specific accuracies for every model, even when their correct predicted probabilities are the lowest. This is primarily driven by TN rather than correct identification, as reflected in its low sensitivity. In this case, this alternative has the lowest balanced accuracy score, compared to a misleadingly high accuracy, highlighting how this metric is more useful for imbalanced settings. These results suggest that models are considerably better at explaining the decision to be vaccinated than at distinguishing between specific vaccine options. It should be stressed that the misclassification we are observing here is at least in part \emph{proper} misclassification rather than just label switching, with changes between free and paid, either within a vaccine or from A to B (and vice versa).

More flexible CM specifications—--such as NL, Asymmetric, Scobit, and Uneven models---show similar patterns. While these models outperform MNL in terms of overall fit, in some cases their predictive performance in the confusion matrices is actually inferior for some alternatives. For example, the NL slightly improves the identification of Free Vaccine B (sensitivity increasing from 0.49 to 0.50 in the testing sample), whereas for the Uneven model its performance decreases (0.47). Similarly, in the training subset, the Scobit model assigns a higher probability to correctly identifying Free Vaccine B relative to the MNL (0.41 versus 0.39). However, it also allocates more probability mass from No Vaccine observations to Free Vaccine B (0.34 versus 0.29), suggesting that the improvement is at the expense of additional misclassifications. The Asymmetric model shows higher TP in the testing set for paid vaccine A (from 0.18 to 0.20), but this is offset by decreases in TP for free vaccine B (from 0.49 to 0.48) and in paid vaccine B (from 0.15 to 0.16). Notably, none of the more flexible specifications substantially improve the identification of the no vaccine alternative in the testing sample, whose sensitivity remains 0.09 across all models in the testing sample. This suggests that the gains in log-likelihood achieved by the more flexible CM structures primarily reflect a redistribution of probabilities among the vaccine alternatives rather than a better identification of the minority alternative. More generally, the confusion matrices reveal how gains and losses are distributed across alternatives, providing a more nuanced understanding of model performance.

Even when turning to ML algorithms, we only observe slight improvements, suggesting a predictive `threshold' regardless of model complexity. This is reinforced by focusing on sensitivity and specificity, where we note that TN classification is excellent, but TP classification performance is very low. Although alternative-specific accuracies appear high, the confusion matrices reveal that a large share of no vaccine observations are misclassified. Consequently, balanced accuracy provides a more informative metric in this imbalanced setting, as it weights sensitivity and specificity equally, preventing the dominant alternative from masking poor performance on the minority alternative, an issue that arises with accuracy, which is heavily influenced by the majority class. 

A pattern emerges when comparing training and testing confusion matrices for the CM specifications. In several cases, out‑of‑sample performance improves in terms of correct classification for specific alternatives. For example, in the NL model, the mean predicted probability for free vaccine B increases substantially from training (0.40) to testing (0.50). Interestingly, this effect is not uniform across alternatives: for paid vaccine B, the opposite pattern is observed, with the mean predicted probability for the correct alternative falling from 0.22 to 0.15 under the NL, alongside similar reductions for free vaccine A (0.42 to 0.36), paid vaccine A (0.25 to 0.18) and no vaccine (0.13 to 0.09). 

An important insight emerges from comparing diagonal elements with market shares. For most alternatives, the mean predicted probabilities are close to their observed frequencies, suggesting that a large portion of predictive performance is driven by alternative‑specific constants rather than by explanatory variables. In other words, the models largely reproduce aggregate market shares but struggle to capture the underlying behavioural mechanisms that distinguish between similar alternatives.

Finally, the confusion matrices show that misclassifications are highly structured rather than random. Misclassification patterns are stronger between similar alternatives---particularly between free and paid versions of the same vaccine---indicating that the models capture broad preference hierarchies but fail to discriminate within closely related options. This further highlights the advantage of probabilistic confusion matrices, as they capture how probabilities are distributed across alternatives, rather than reducing predictions to a single deterministic outcome.

\begin{table}[h]
\centering
\scriptsize
\renewcommand{\arraystretch}{1.15}
\caption{Log-likelihood and ML metrics for the Covid Vaccine dataset.}
\begin{tabular}{l|*{5}{c}|*{5}{c}}
\multirow{2}{*}{}
  & \multicolumn{5}{c|}{\textbf{Training}}
  & \multicolumn{5}{c}{\textbf{Testing}} \\
\cline{2-11}
  & \textbf{LL} & \textbf{Acc} & \textbf{Sens} & \textbf{Spec} & \textbf{Bal}
  & \textbf{LL} & \textbf{Acc} & \textbf{Sens} & \textbf{Spec} & \textbf{Bal} \\
\hline
\multicolumn{1}{l|}{\textbf{Multinomial logit}} &  -13,322.35 & 0.33 &  &  &  & -3,468.08 & 0.34 &  &  &  \\
Free Vaccine A & & 0.62 & 0.41 & 0.72 & 0.56 & & 0.78 & 0.36 & 0.85 & 0.61 \\
Paid Vaccine A & & 0.75 & 0.25 & 0.86 & 0.55 & & 0.81 & 0.18 & 0.91 & 0.54 \\
Free Vaccine B & & 0.63 & 0.39 & 0.74 & 0.57 & & 0.57 & 0.49 & 0.62 & 0.56 \\
Paid Vaccine B & & 0.77 & 0.22 & 0.86 & 0.54 & & 0.78 & 0.15 & 0.89 & 0.52 \\
No Vaccine & & 0.88 & 0.15 & 0.94 & 0.54 & & 0.90 & 0.09 & 0.95 & 0.52 \\
\hline
\multicolumn{1}{l|}{\textbf{Nested logit}} & -13,227.76 & 0.34 &  &  &  & -3,449.042 & 0.34 &  &  &  \\
Free Vaccine A & & 0.63 & 0.42 & 0.72 & 0.57 & & 0.78 & 0.36 & 0.86 & 0.61 \\
Paid Vaccine A & & 0.75 & 0.25 & 0.86 & 0.55 & & 0.81 & 0.18 & 0.91 & 0.54 \\
Free Vaccine B & & 0.64 & 0.40 & 0.75 & 0.57 & & 0.57 & 0.50 & 0.62 & 0.56 \\
Paid Vaccine B & & 0.78 & 0.22 & 0.86 & 0.54 & & 0.79 & 0.15 & 0.89 & 0.52 \\
No Vaccine & & 0.88 & 0.13 & 0.94 & 0.53 & & 0.89 & 0.09 & 0.95 & 0.52 \\
\hline
\multicolumn{1}{l|}{\textbf{Asymmetric logit}} & -13,129.65 & 0.35 &  &  &  & -3,425.06 & 0.34 &  &  & \\
Free Vaccine A & & 0.64 & 0.42 & 0.73 & 0.58 & & 0.79 & 0.35 & 0.87 & 0.61 \\
Paid Vaccine A & & 0.75 & 0.27 & 0.85 & 0.56 & & 0.80 & 0.20 & 0.89 & 0.55 \\
Free Vaccine B & & 0.64 & 0.42 & 0.74 & 0.58 & & 0.57 & 0.48 & 0.63 & 0.56 \\
Paid Vaccine B & & 0.79 & 0.22 & 0.88 & 0.55 & & 0.78 & 0.16 & 0.89 & 0.52 \\
No Vaccine & & 0.88 & 0.13 & 0.94 & 0.53 & & 0.89 & 0.09 & 0.94 & 0.52 \\
\hline
\multicolumn{1}{l|}{\textbf{Multinomial scobit}} & -13,175.8 & 0.34 &  &  &  & -3,435.29 & 0.34 &  &  &  \\
Free Vaccine A & & 0.63 & 0.41 & 0.73 & 0.57 & & 0.79 & 0.35 & 0.86 & 0.61 \\
Paid Vaccine A & & 0.75 & 0.27 & 0.85 & 0.56 & & 0.81 & 0.20 & 0.90 & 0.55 \\
Free Vaccine B & & 0.64 & 0.41 & 0.74 & 0.57 & & 0.57 & 0.50 & 0.62 & 0.56 \\
Paid Vaccine B & & 0.78 & 0.21 & 0.87 & 0.54 & & 0.79 & 0.15 & 0.89 & 0.52 \\
No Vaccine & & 0.88 & 0.13 & 0.94 & 0.53 & & 0.89 & 0.09 & 0.95 & 0.52 \\
\hline
\multicolumn{1}{l|}{\textbf{Uneven logit}} & -13,170.46 & 0.34 &  &  &  & -3,435.23 & 0.34 &  &  &  \\
Free Vaccine A & & 0.63 & 0.43 & 0.72 & 0.58 & & 0.79 & 0.36 & 0.86 & 0.61 \\
Paid Vaccine A & & 0.75 & 0.25 & 0.86 & 0.55 & & 0.80 & 0.19 & 0.90 & 0.54 \\
Free Vaccine B & & 0.64 & 0.40 & 0.75 & 0.57 & & 0.57 & 0.47 & 0.64 & 0.55 \\
Paid Vaccine B & & 0.78 & 0.23 & 0.87 & 0.55 & & 0.78 & 0.17 & 0.88 & 0.52 \\
No Vaccine & & 0.88 & 0.13 & 0.94 & 0.53 & & 0.89 & 0.09 & 0.94 & 0.52 \\
\hline
\multicolumn{1}{l|}{\textbf{Random forest}} & -10,591.78 & 0.36 &  &  &  & -3,438.05 & 0.33 &  &  &  \\
Free Vaccine A & & 0.64 & 0.43 & 0.74 & 0.58 & & 0.78 & 0.42 & 0.86 & 0.64 \\
Paid Vaccine A & & 0.76 & 0.30 & 0.85 & 0.58 & & 0.75 & 0.28 & 0.85 & 0.56 \\
Free Vaccine B & & 0.65 & 0.43 & 0.74 & 0.59 & & 0.62 & 0.38 & 0.74 & 0.56 \\
Paid Vaccine B & & 0.79 & 0.23 & 0.88 & 0.55 & & 0.78 & 0.22 & 0.87 & 0.54 \\
No Vaccine & & 0.88 & 0.13 & 0.94 & 0.53 & & 0.87 & 0.10 & 0.94 & 0.52 \\
\hline
\multicolumn{1}{l|}{\textbf{Artificial neural network}} & -9,076.75 & 0.34 &  &  &  & -3,388.34 & 0.33 &  &  &  \\
Free Vaccine A & & 0.64 & 0.40 & 0.74 & 0.57 & & 0.79 & 0.40 & 0.87 & 0.63 \\
Paid Vaccine A & & 0.77 & 0.26 & 0.86 & 0.56 & & 0.76 & 0.25 & 0.86 & 0.56 \\
Free Vaccine B & & 0.62 & 0.43 & 0.71 & 0.57 & & 0.62 & 0.43 & 0.71 & 0.57 \\
Paid Vaccine B & & 0.78 & 0.20 & 0.87 & 0.53 & & 0.78 & 0.19 & 0.87 & 0.53 \\
No Vaccine & & 0.89 & 0.11 & 0.94 & 0.52 & & 0.87 & 0.09 & 0.94 & 0.52 \\
\hline
\multicolumn{1}{l|}{\textbf{XGBoost}} & -8,955.02 & 0.35 &  &  &  & -3,364.17 & 0.34 &  &  &  \\
Free Vaccine A & & 0.64 & 0.40 & 0.74 & 0.57 & & 0.79 & 0.41 & 0.87 & 0.64 \\
Paid Vaccine A & & 0.77 & 0.27 & 0.86 & 0.57 & & 0.76 & 0.26 & 0.86 & 0.56 \\
Free Vaccine B & & 0.62 & 0.43 & 0.72 & 0.58 & & 0.62 & 0.43 & 0.72 & 0.57 \\
Paid Vaccine B & & 0.78 & 0.22 & 0.87 & 0.54 & & 0.78 & 0.21 & 0.87 & 0.54 \\
No Vaccine & & 0.89 & 0.13 & 0.94 & 0.53 & & 0.87 & 0.10 & 0.94 & 0.52 \\

\end{tabular}
\label{tab:metrics_covid}
\end{table}

\newgeometry{margin=1.5cm}
\begin{landscape}
\begin{figure}[p]
\centering
\cmpanel{covidsubfig}{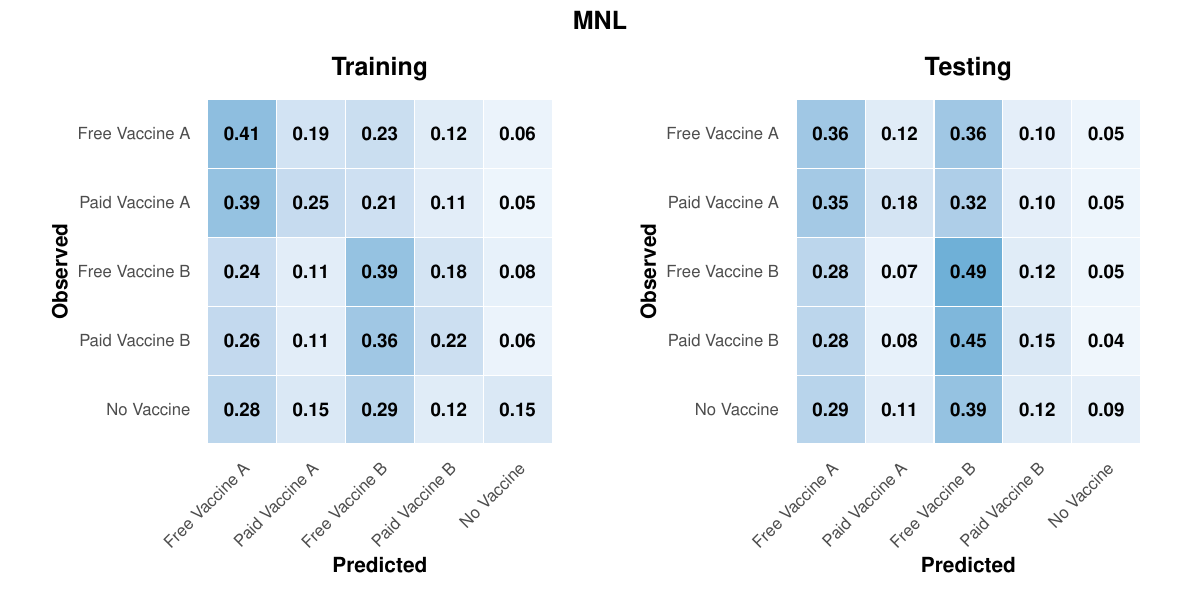}{MNL}{Covid}{fig:covid_2A}\hfill
\cmpanel{covidsubfig}{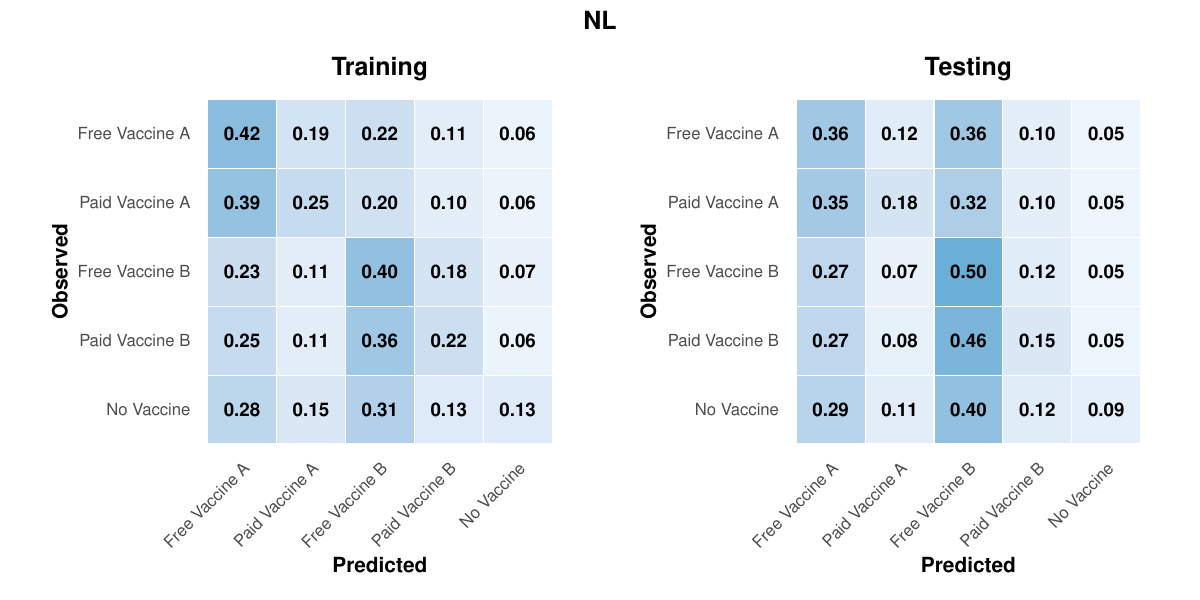}{NL}{Covid}{fig:covid_2B}\hfill
\cmpanel{covidsubfig}{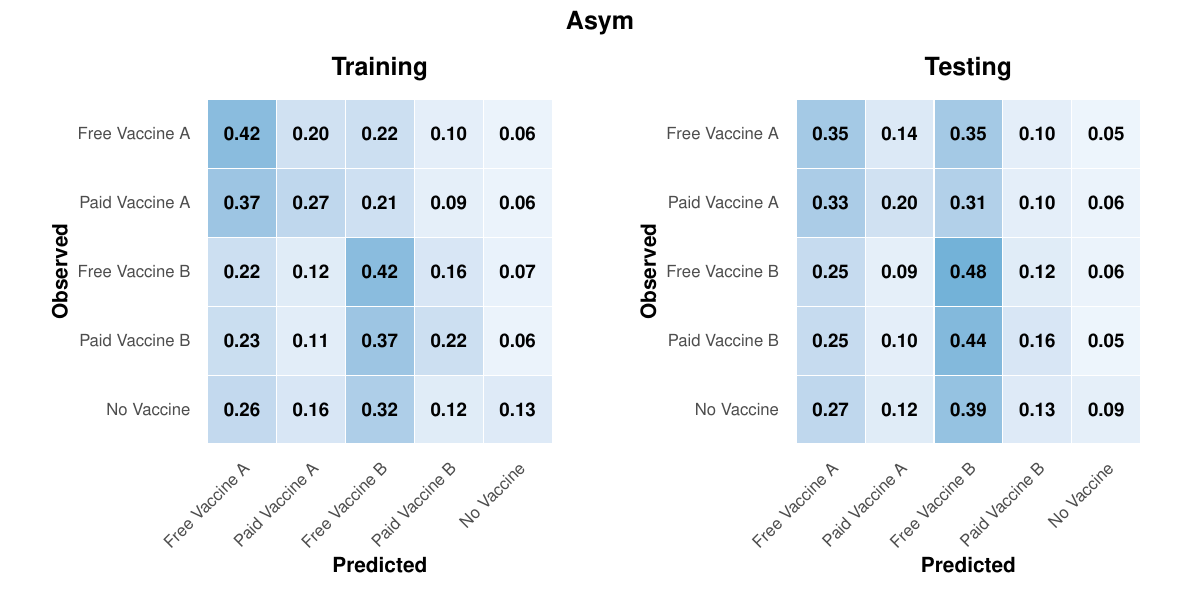}{Asymmetric}{Covid}{fig:covid_2C}\hfill
\cmpanel{covidsubfig}{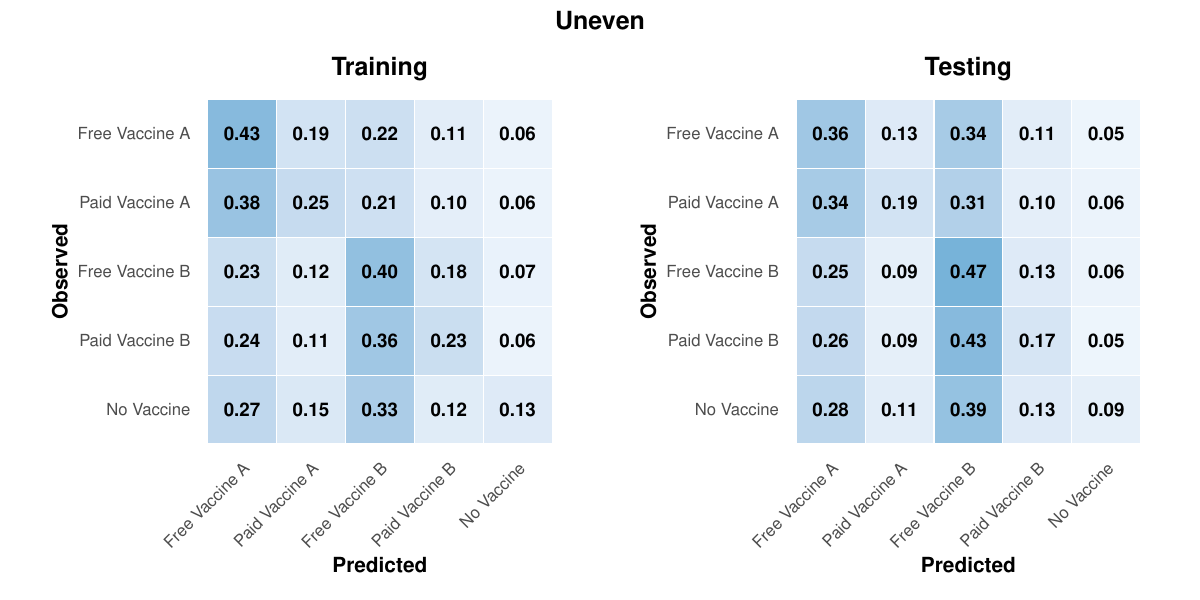}{Uneven}{Covid}{fig:covid_2D}

\vspace{1.5em}

\cmpanel{covidsubfig}{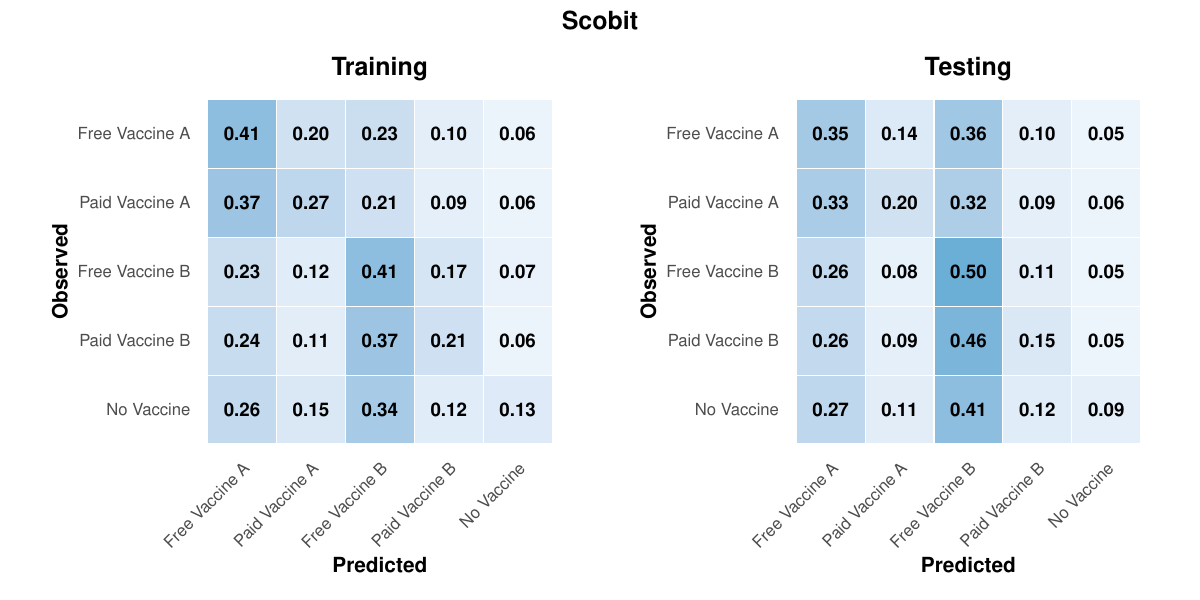}{Scobit}{Covid}{fig:covid_2E}\hfill
\cmpanel{covidsubfig}{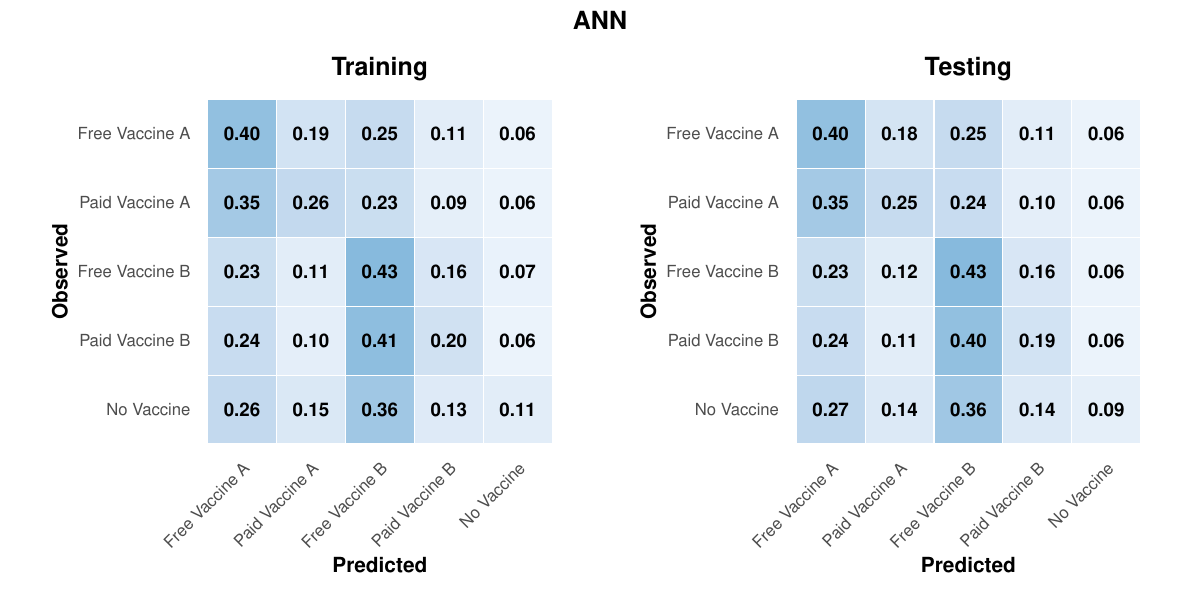}{ANN}{Covid}{fig:covid_2F}\hfill
\cmpanel{covidsubfig}{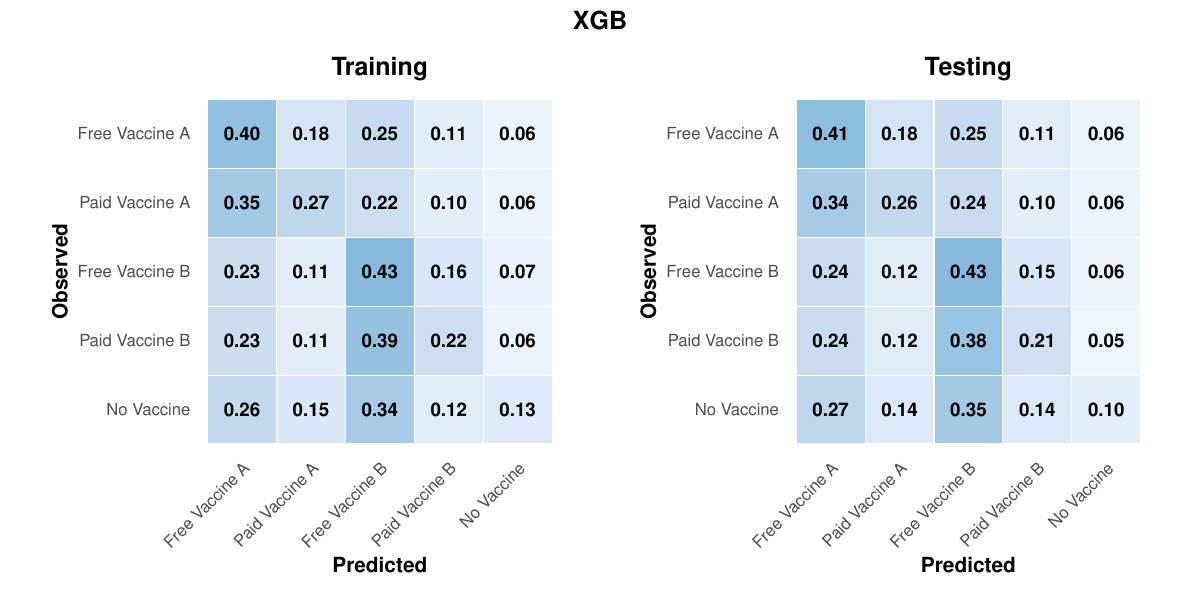}{XGB}{Covid}{fig:covid_2G}\hfill
\cmpanel{covidsubfig}{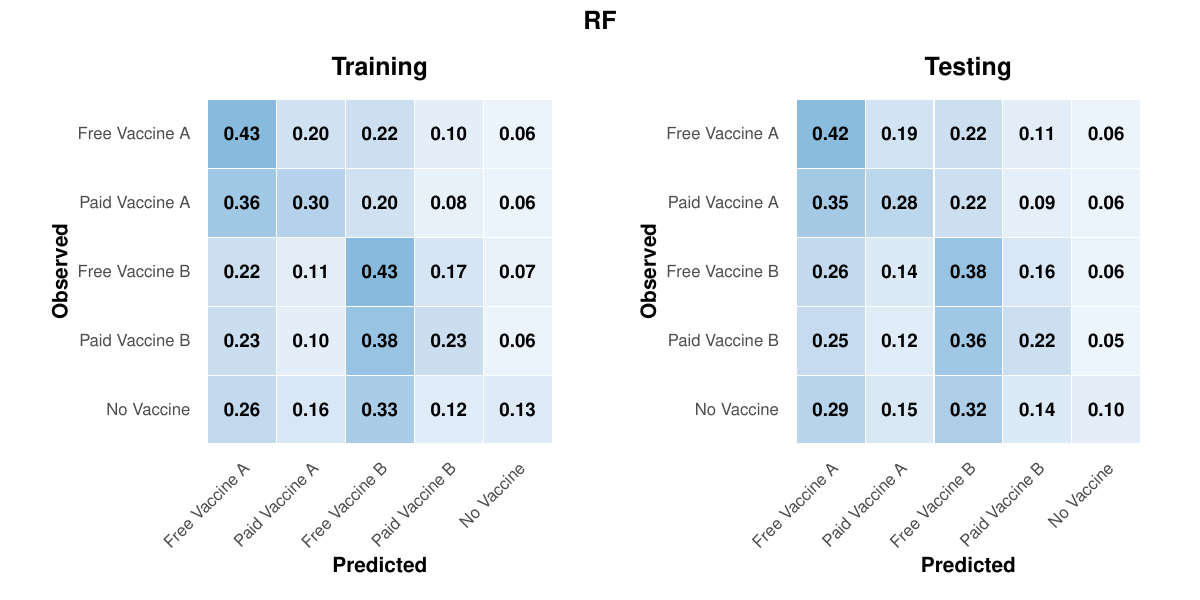}{RF}{Covid}{fig:covid_2H}
\end{figure}
\end{landscape}
\restoregeometry

\subsection{SwissMetro dataset}
\label{subsec:case_swiss}

The SwissMetro dataset originates from an SP survey collected on trains between St. Gallen and Geneva, Switzerland, and has been widely used in the literature \citep{sifringer2020, salas2025}. This dataset considers three alternatives: Train, Car (only available to car owners), and a proposed high-speed public transport system, SwissMetro. The original dataset contained 10,728 observations, and we removed respondents holding a season ticket (who face zero monetary cost), yielding 9,207 observations. SwissMetro is the most frequently chosen alternative (58.5\%), followed by Car (32.8\%) and Train (8.7\%). Available attributes included are travel time, travel cost, frequency, age, luggage and the type of seat. A detailed description can be found in \cite{swissmetro}.

We report overall fit using log-likelihood (LL), and provide alternative-specific metrics using accuracy, sensitivity, specificity, and balanced accuracy, in \autoref{tab:metrics_swissmetro}. We also present the confusion matrices for this dataset from \autoref{fig:swiss_3A} to \autoref{fig:swiss_3H}. 

The CM specifications exhibit very similar performance, with slight improvements from more flexible structures, such as the NL or the Scobit formulations. Therefore, the unobserved differences captured by these more complex structures are very small in terms of the predicted probabilities. In contrast, ML models achieve substantial better fit in both the training and testing data, particularly RF and XGB. However, for these models, the performance in training data does not necessarily translate into superior out-of-sample fit, in terms of the ML metrics. This difference between in-sample and testing behaviour does not indicate overfitting in this case---our validation checks confirm that ML models generalise well---but rather highlights that a high training log-likelihood does not necessarily translate into superior out-of-sample performance when evaluated using alternative-specific metrics.

Across all models---both CM and ML---confusion matrices show a relatively consistent structure. SwissMetro is the dominant alternative, with the highest mean probability in every model. Train is the most challenging alternative: the majority of Train observations are predicted as SwissMetro, indicating substantial misclassification patterns between these two rail-based modes. Furthermore, for Train, misclassifications are higher than correct classifications for all cases. This pattern highlights where modelling effort should be concentrated. Rather than pursuing incremental improvements in aggregate fit, the confusion matrix suggests that the greatest gains are likely to come from better explaining the behavioural differences between Train and SwissMetro. This information is not available from log-likelihood alone, which indicates overall fit but does not reveal which alternatives are driving misclassifications.

Most of the literature indicates that ML outperforms CM in deterministic outputs. In terms of overall fit, our results confirm this for probabilistic outputs, as all ML models achieve superior log-likelihood values in both training and testing. However, when evaluating probabilistic ML metrics, the performance of CM and ML models is remarkably similar, with CM models occasionally matching or slightly outperforming the ANN across specific alternative-level metrics. 

Across all models, most alternatives exhibit accuracy levels above the 70\% threshold often used as an acceptable benchmark in the literature. This highlights that accuracy itself is not sufficient for analysing model performance, let alone deterministic accuracy. SwissMetro consistently achieves high sensitivity, whereas Train remains the weakest alternative for every specification. Car displays intermediate scores for both sensitivity and specificity, and this alternative's predictions are more stable and less affected by model structure. Regardless of the model, we obtain a substantial share of misclassifications. We emphasise sensitivity because CM evaluation is traditionally based on TP (for computing log-likelihood).

These results are consistent with the patterns observed in the confusion matrices and highlight an important distinction between CM and ML. Because CM is theory-driven and often recover market shares, the training and testing confusion matrices are very similar, as the probability allocation across alternatives is largely driven by model structure.  In contrast, ML algorithms are highly flexible and data-driven, and therefore the distribution of predicted probabilities can differ more between training and testing samples. These differences do not necessarily imply overfitting but rather reflect the fact that ML models do not enforce behavioural constraints such as market‑share recovery or alternative‑specific utilities. As a result, an ML model may achieve excellent fit in the training data yet show weaker separation between alternatives in the test data. Although one could regularise or restrict ML models to reduce this gap, doing so typically comes at the expense of other desirable properties, illustrating the trade‑off between flexibility and generalisability.

Indeed, ANN, XGB, and RF achieve very high in-sample sensitivity. However, even the best performing model (XGB) exhibits a significant sensitivity gap between training and testing splits, and between high-share and low-share alternatives. This reflects the challenges of predicting minority alternatives. However, before settling on the final ML model specifications, we tested several ANN, XGB, and RF architectures, and frequently observed that their training confusion matrices differed considerably from the testing confusion matrices, particularly for the minority alternative: Train. This initially suggested overfitting, as these models yielded higher sensitivities for the training split than for the testing split. For all the models, we conducted a formal overfitting check by comparing both the confusion matrices and the validation vs training losses, and we show the results for the final specifications in \autoref{appendix:overfitting}. This reinforces the use of the confusion matrix as a diagnostic tool for detecting overfitting.

Specificity results show an additional limitation: CM structures struggle to correctly identify TN for SwissMetro, meaning that this alternative receives high predicted probabilities even when it is not the observed choice. In contrast, specificity for Car and Train is more stable. Finally, the balanced accuracy scores reveal weaknesses that accuracy fails to show: in the SwissMetro dataset, accuracy remains high even when models rarely identify the low‑share alternatives, whereas balanced accuracy drops sharply because it reflects the very low specificity for SwissMetro and the uneven sensitivity across alternatives.

Moreover, two models may obtain almost identical aggregate metrics---such as accuracy or log-likelihood---while having remarkably different confusion matrices. This could happen because both accuracy and log-likelihood focus on correct predictions aggregated across all alternatives, without revealing how probabilities are distributed among misclassifications. For example, Scobit and XGB have a similar score on overall accuracy (0.604 and 0.605, respectively), but their confusion matrices are considerably different, when looking at the differences between Train TP and FN, and Car FN. In the testing sample, Train shows a clear difference between the two models: Scobit correctly identifies 22\% of Train observations, whereas XGB reaches 29\%, a 31.8\% relative improvement. This can also be seen in FN: under Scobit, 62\% of Train observations are misclassified as SwissMetro, while under XGB this falls to 56\%.

\begin{table}[h]
\centering
\scriptsize
\renewcommand{\arraystretch}{1.15}
\caption{Log-likelihood and ML metrics for the SwissMetro dataset.}
\begin{tabular}{l|*{5}{c}|*{5}{c}}
\multirow{2}{*}{}
  & \multicolumn{5}{c|}{\textbf{Training}}
  & \multicolumn{5}{c}{\textbf{Testing}} \\
\cline{2-11}
  & \textbf{LL} & \textbf{Acc} & \textbf{Sens} & \textbf{Spec} & \textbf{Bal}
  & \textbf{LL} & \textbf{Acc} & \textbf{Sens} & \textbf{Spec} & \textbf{Bal} \\
\hline
\multicolumn{1}{l|}{\textbf{Multinomial logit}} & -4,756.67 & 0.61 & & & & -1,246.01& 0.60& & & \\
Train       & & 0.88 & 0.27 & 0.93 & 0.60 & &  0.87  & 0.22 & 0.93 & 0.58 \\
SwissMetro  & & 0.64 & 0.69 & 0.56 & 0.62 & &  0.63  & 0.68 & 0.56 & 0.62 \\
Car         & & 0.71 & 0.56 & 0.78 & 0.67 & &  0.71  & 0.57 & 0.78 & 0.67 \\
\hline
\multicolumn{1}{l|}{\textbf{Nested logit}} & -4,755.38 & 0.61& & & & -1,244.83 &0.60 & \\
Train       & & 0.88 & 0.27 & 0.93 & 0.60 & & 0.87 & 0.23 & 0.93 & 0.58 \\
SwissMetro  & & 0.64 & 0.69 & 0.56 & 0.62 & & 0.63 & 0.68 & 0.56 & 0.62 \\
Car         & & 0.71 & 0.56 & 0.78 & 0.67 & & 0.71 & 0.57 & 0.78 & 0.67 \\
\hline
\multicolumn{1}{l|}{\textbf{Asymmetric logit}} & -4,749.87&0.61 & & & &-1,244.35 & 0.60& \\
Train       & & 0.88 & 0.27 & 0.93 & 0.60 & & 0.87 & 0.22 & 0.93 & 0.58 \\
SwissMetro  & & 0.64 & 0.69 & 0.56 & 0.63 & & 0.63 & 0.68 & 0.56 & 0.62 \\
Car         & & 0.71 & 0.56 & 0.78 & 0.67 & & 0.71 & 0.57 & 0.78 & 0.67 \\
\hline
\multicolumn{1}{l|}{\textbf{Multinomial scobit}} & -4,753.57& 0.61& & & &-1,245.37 &0.60 & \\
Train       & & 0.88 & 0.27 & 0.93 & 0.60 & & 0.87 & 0.22 & 0.93 & 0.58 \\
SwissMetro  & & 0.64 & 0.69 & 0.56 & 0.62 & & 0.63 & 0.68 & 0.56 & 0.62 \\
Car         & & 0.71 & 0.56 & 0.78 & 0.67 & & 0.71 & 0.57 & 0.78 & 0.67 \\
\hline
\multicolumn{1}{l|}{\textbf{Uneven logit}} & -4,747.50 &0.61 & & & &-1,244.43 &0.61 & \\
Train       & & 0.88 & 0.27 & 0.93 & 0.60 &  & 0.87 & 0.23 & 0.93 & 0.58 \\
SwissMetro  & & 0.64 & 0.69 & 0.56 & 0.63 &  & 0.63 & 0.68 & 0.56 & 0.62 \\
Car         & & 0.71 & 0.56 & 0.78 & 0.67 &  & 0.71 & 0.57 & 0.78 & 0.67 \\
\hline

\multicolumn{1}{l|}{\textbf{Random forest}} &-2,937.32 &0.69 & & & & -1,072.98& 0.64& \\
Train       & & 0.90 & 0.41 & 0.95 & 0.68 & & 0.88 & 0.29 & 0.94 & 0.61 \\
SwissMetro  & & 0.71 & 0.76 & 0.66 & 0.71 & & 0.66 & 0.71 & 0.59 & 0.65 \\
Car         & & 0.77 & 0.66 & 0.83 & 0.74 & & 0.73 & 0.60 & 0.80 & 0.70 \\
\hline
\multicolumn{1}{l|}{\textbf{Artificial neural network}} &-4,316.19 & 0.60& & & &-1,233.77 & 0.59& \\
Train       & & 0.87 & 0.24 & 0.93 & 0.58 & & 0.86 & 0.21 & 0.93 & 0.57 \\
SwissMetro  & & 0.62 & 0.67 & 0.55 & 0.61 & & 0.62 & 0.67 & 0.55 & 0.61 \\
Car         & & 0.70 & 0.55 & 0.77 & 0.66 & & 0.70 & 0.56 & 0.77 & 0.66 \\
\hline
\multicolumn{1}{l|}{\textbf{XGBoost}} &-3,908.32 &0.61 & & & & -1,155.79& 0.61& \\
Train       & & 0.88 & 0.29 & 0.93 & 0.61 & & 0.87 & 0.29 & 0.93 & 0.61 \\
SwissMetro  & & 0.64 & 0.69 & 0.57 & 0.65 & & 0.63 & 0.69 & 0.56 & 0.62 \\
Car         & & 0.71 & 0.56 & 0.78 & 0.67 & & 0.70 & 0.55 & 0.78 & 0.66 \\

\end{tabular}
\label{tab:metrics_swissmetro}
\end{table}

\newgeometry{margin=1.5cm}
\begin{landscape}
\begin{figure}[p]
\centering
\cmpanel{swisssubfig}{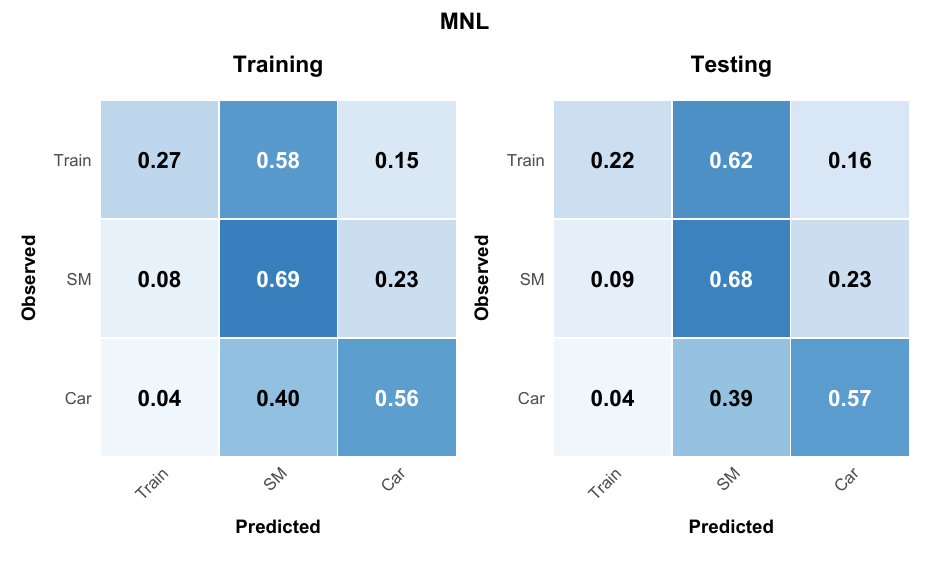}{MNL}{SwissMetro}{fig:swiss_3A}\hfill
\cmpanel{swisssubfig}{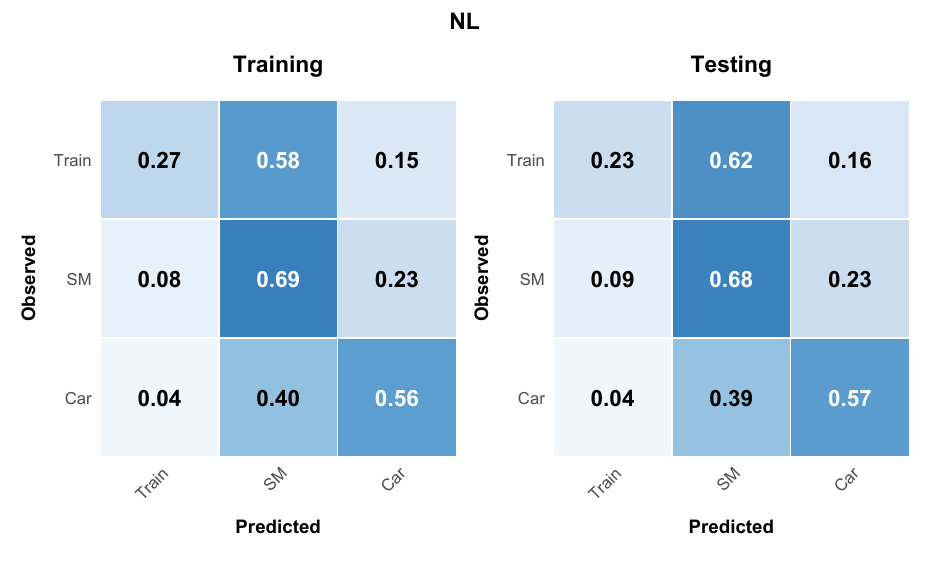}{NL}{SwissMetro}{fig:swiss_3B}\hfill
\cmpanel{swisssubfig}{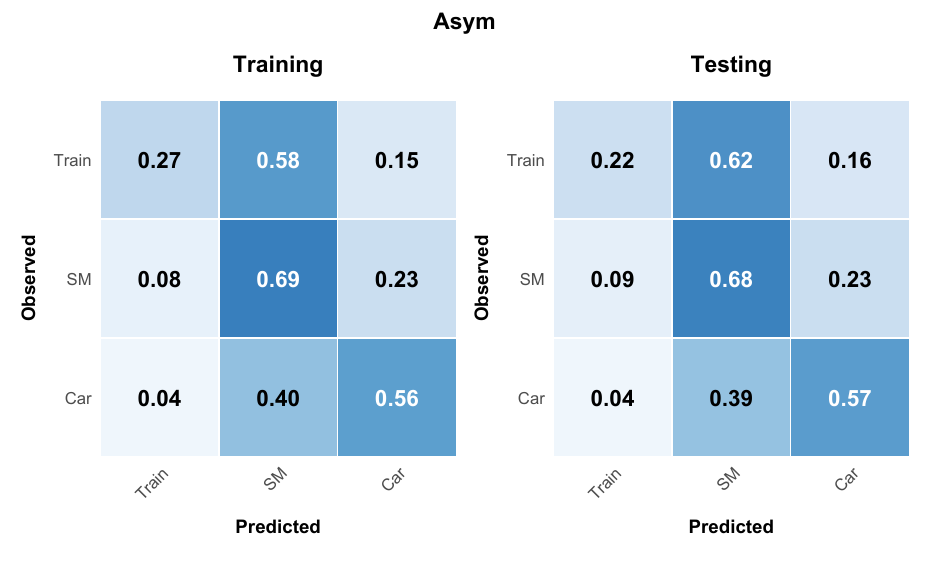}{Asymmetric}{SwissMetro}{fig:swiss_3C}\hfill
\cmpanel{swisssubfig}{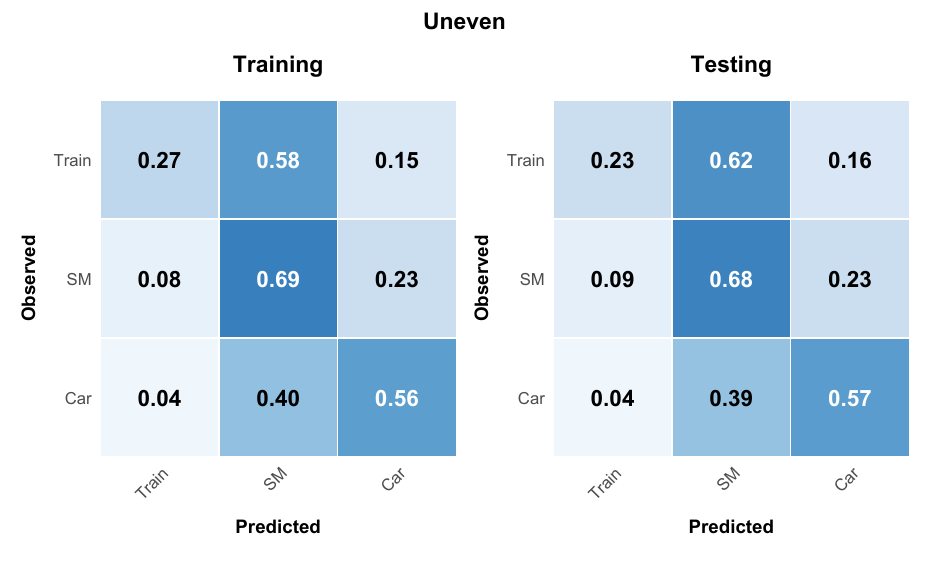}{Uneven}{SwissMetro}{fig:swiss_3D}

\vspace{1.5em}

\cmpanel{swisssubfig}{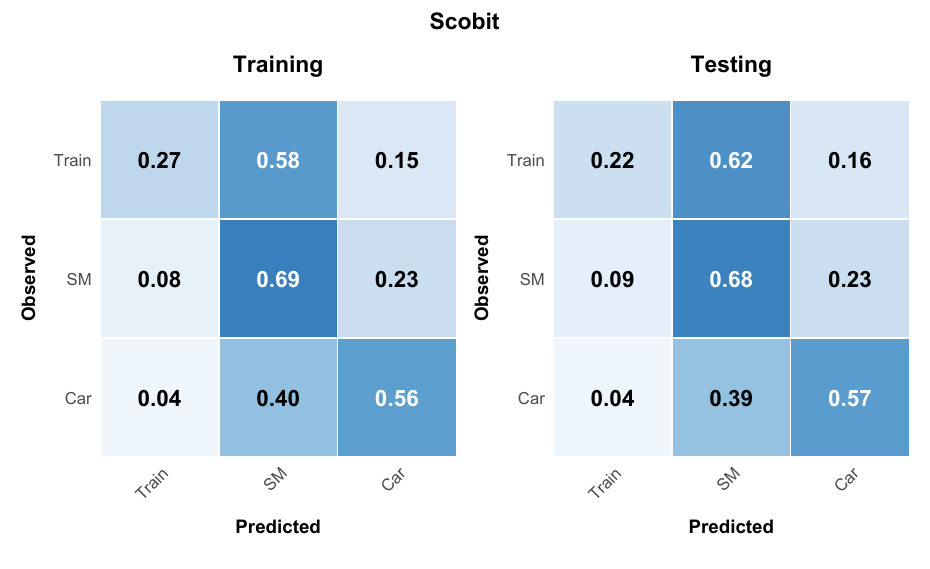}{Scobit}{SwissMetro}{fig:swiss_3E}\hfill
\cmpanel{swisssubfig}{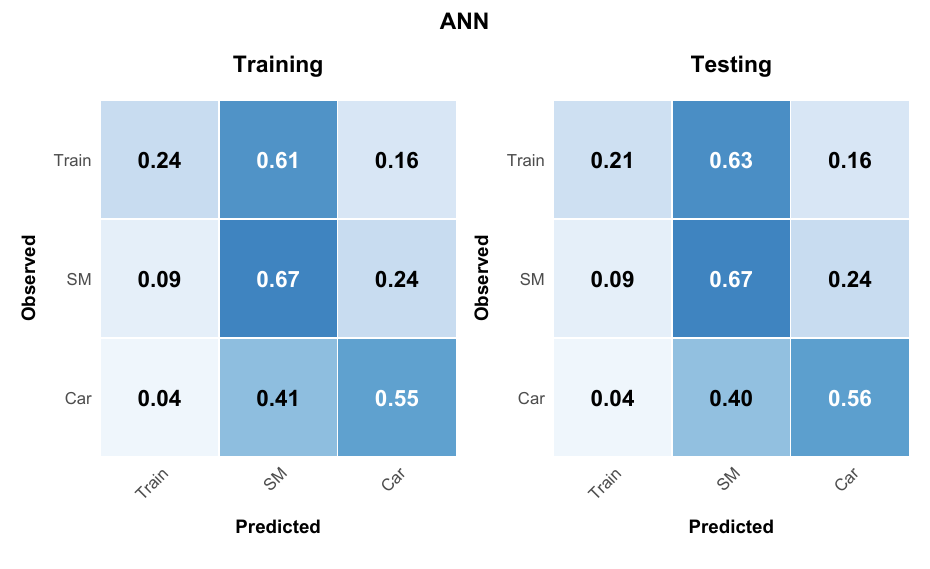}{ANN}{SwissMetro}{fig:swiss_3F}\hfill
\cmpanel{swisssubfig}{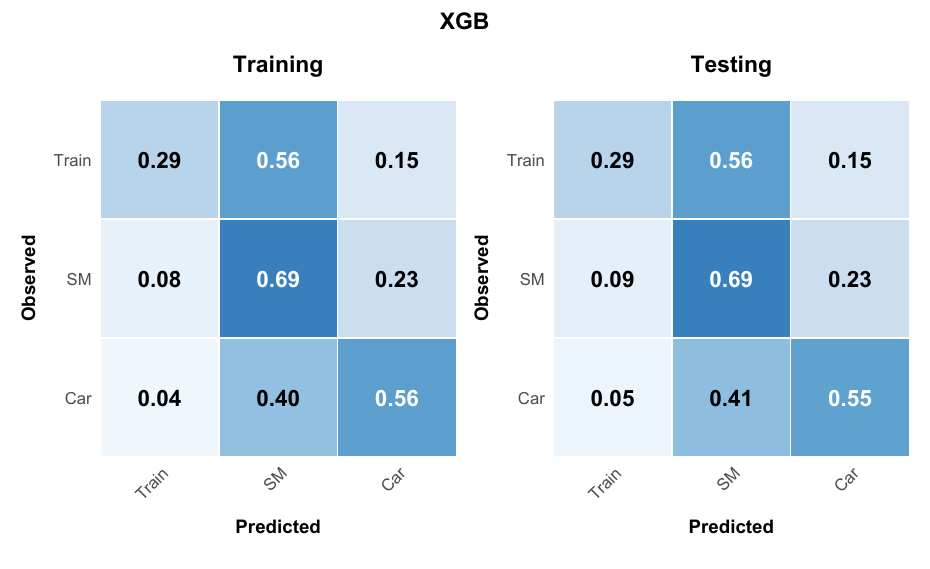}{XGB}{SwissMetro}{fig:swiss_3G}\hfill
\cmpanel{swisssubfig}{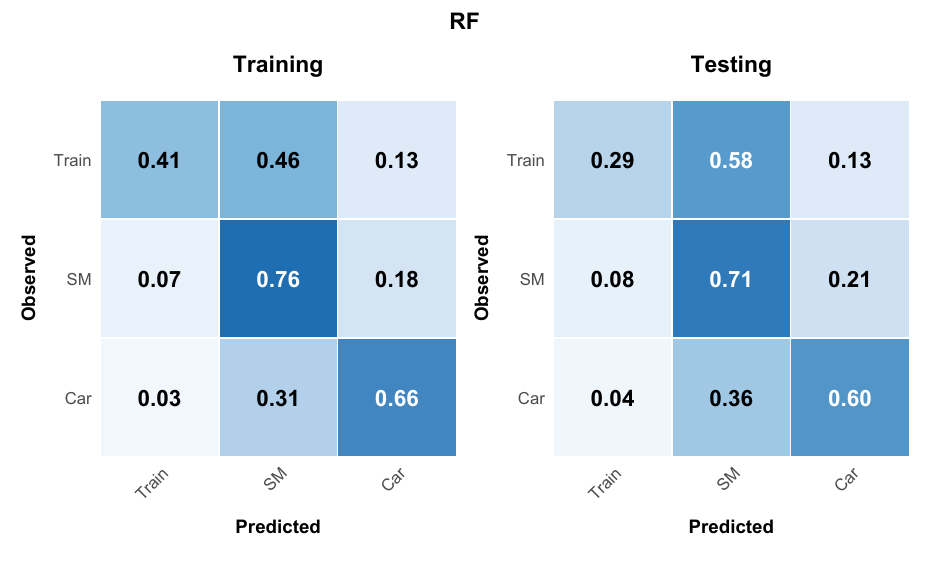}{RF}{SwissMetro}{fig:swiss_3H}
\end{figure}
\end{landscape}
\restoregeometry

\subsection{Decisions dataset}
\label{subsec:case_decisions}

The Decisions dataset was collected by the Choice Modelling Centre (University of Leeds) between November 2016 and April 2017 as part of a large-scale survey on life and travel patterns, activities, and social networks. After data cleaning, the sample comprises 12,004 trips from 415 individuals, with six alternatives: Car (48.1\%), Bus (14.4\%), Rail (5.0\%), Taxi (3.3\%), Bicycle (3.3\%), and Walking (26.0\%). A detailed description is presented by \cite{decisions}. 

Analogous to the previous case studies, we report overall fit and the full set of ML metrics in \autoref{tab:metrics_decisions}, and the confusion matrices from \autoref{fig:dec_4A} to \autoref{fig:dec_4H}. In this dataset, overall performance is superior compared to the SwissMetro case study, for both CM and ML algorithms. This can be concluded from the structure of the confusion matrices: the diagonal elements are considerably higher, and the off-diagonal entries are relatively small. This improvement suggests that predictive performance depends not only on choice imbalance, but also on how easily alternatives can be distinguished from one another. Although the Decisions dataset is highly imbalanced, its alternatives exhibit greater differences between attributes than those in the Covid dataset, allowing models to more easily distinguish observed choices based on the explanatory variables.

Indeed, we can see a clear pattern across all models: the majority alternatives---Car and Walking---show the highest probabilities for correct predictions, and the lowest probabilities for incorrect predictions. In contrast, the minority alternatives---Taxi and Cycling---show the lowest probabilities in the diagonal, and the highest probabilities in the off-diagonal elements. It is also striking how Bus receives a substantial share of predicted probabilities for Taxi and Cycling observations, particularly in the testing sample. This reveals misclassification patterns between Bus and the low-share alternatives, suggesting that Taxi and Cycling journeys occur in contexts where Bus is also a feasible choice (e.g., short trips). Hence, for targeting improvements in model fit, we could focus on refining the utility function of Bus, for instance by incorporating more covariates, non-linearities on some attributes, introducing interactions, or changing model structure. This shows direct implications of the use of these metrics in model development.

For this case study, we note larger differences between models compared to the other datasets, both in log-likelihood and in the confusion matrices. Among the CM structures, the Scobit model delivers the greatest improvements over MNL, driven primarily by increases in Rail sensitivity (by 4\%), while decreasing Taxi and Cycling sensitivities by 1\% each. Likewise, the Uneven model outperforms MNL in terms of overall fit, but in this case, the improvement is driven by a 2\% increase in Taxi sensitivity. 

We also analyse how aggregate metrics can mask alternative-specific performance differences. For example, MNL and XGB achieve very similar out-of-sample log-likelihoods (1,149.87 and 1,149.08, respectively), yet their corresponding confusion matrices differ substantially. This can be seen for Car, Rail, and Taxi in the diagonal elements (see \autoref{fig:dec_4A} and \autoref{fig:dec_4G}). This also suggests that two models with nearly identical accuracy and very similar log-likelihood can allocate probabilities in very different ways across alternatives, and therefore have different confusion matrices. These distinctions are also behaviourally meaningful, as they can affect economic indicators derived from CM, such as marginal rates of substitution and elasticities. 

Other CM structures show different trade-offs. For example, the Asymmetric model shows a decrease in Cycling TP but compensates with improved Car, Rail, and Taxi TP, indicating a reallocation of probability that, in this case, benefits the log-likelihood. The NL performs worse than the MNL on out-of-sample log-likelihood, despite achieving equal or higher accuracy across all alternatives. This highlights another important finding: the relationship between training and testing log-likelihood is not monotonic, i.e., even when training fit improves, testing fit can decrease. 

Accuracy scores are excellent across all models, splits, and alternatives. However, these must be interpreted with caution. As we mentioned before, accuracy may be biased in imbalanced settings, particularly in multi-alternative cases. For example, in the MNL testing split, Taxi achieves an accuracy of 0.89, but a sensitivity of only 0.27. This indicates that this accuracy score is largely driven by TN, not the correct predictions of Taxi trips. In other words, the model achieves a high accuracy for this alternative because it is rarely chosen. This also applies to ML algorithms, where accuracy is the most used metric in the literature, and some alternatives achieve an almost perfect accuracy. 

Sensitivity shows that the majority alternatives are exceptionally well predicted, whereas rarely chosen alternatives have lower scores for this metric. Specificity, on the other hand, is extremely high for the low-share alternatives. The combination of low sensitivity and high specificity highlights the severe imbalance of the dataset, and how models tend to prioritise alternatives with a higher market share. Balanced accuracy is particularly informative in this context, as it penalises models that perform well on the majority alternatives but poorly on the minority ones, making it more informative than accuracy for some cases. This metric also shows that, while CM achieve moderate performance for the minority alternatives, ML models substantially improve these scores. 

For all models, Cycling shows the highest differences between training and testing matrices. This could be attributed to Cycling trips occurring under a highly context-dependent set of conditions---such as weather, topography, and cycle lane availability---making these trips harder to predict. This demonstrates that rarely chosen alternatives are the first to lose predictive power when models are evaluated out-of-sample, an insight that becomes clearer through a probabilistic confusion matrix. Also, although Taxi and Bicycle have almost identical market shares, their predictive metrics differ noticeably. This highlights that market share alone does not determine how well an alternative can be predicted, and could be explained because taxi choices are more uniform, whereas bicycle is mostly available only for short trips. Indeed, rarely chosen alternatives are badly predicted when the differences between alternatives are not clear. For instance, Rail trips are often characterised by distinct travel patterns and service attributes that differentiate them from other modes, making them easier to identify despite their relatively low market share. Conversely, the distinction between Car and Taxi trips is often less clear, as these modes may occur in similar travel contexts and share comparable characteristics, making them harder to predict correctly.

Similar to the SwissMetro dataset, our initial experiments with different ANN, XGB, and RF configurations revealed even larger gaps between the training and testing confusion matrices, which initially suggested possible overfitting. We therefore used the confusion matrix as a diagnostic tool to refine model specifications, and we present the overfitting check for the final specifications in \autoref{appendix:overfitting}, where we confirm that they generalise well. 

\begin{table}[h]
\centering
\scriptsize
\renewcommand{\arraystretch}{1.15}
\caption{ML metrics for the Decisions dataset.}
\begin{tabular}{l|*{5}{c}|*{5}{c}}
\multirow{2}{*}{}
  & \multicolumn{5}{c|}{\textbf{Training}}
  & \multicolumn{5}{c}{\textbf{Testing}} \\
\cline{2-11}
  & \textbf{LL} &\textbf{Acc} & \textbf{Sens} & \textbf{Spec} & \textbf{Bal}
  & \textbf{LL} &\textbf{Acc} & \textbf{Sens} & \textbf{Spec} & \textbf{Bal} \\
\hline
\multicolumn{1}{l|}{\textbf{Multinomial logit}} & -3,292.91 & 0.82 & & & & -1,149.87& 0.81& \\
Car       & & 0.94 & 0.93 & 0.94 & 0.94 & & 0.92 & 0.92 & 0.94 & 0.93 \\
Bus       & & 0.90 & 0.63 & 0.94 & 0.79 & & 0.88 & 0.65 & 0.94 & 0.79 \\
Rail      & & 0.96 & 0.67 & 0.98 & 0.83 & & 0.94 & 0.63 & 0.97 & 0.80 \\
Taxi      & & 0.90 & 0.28 & 0.98 & 0.63 & & 0.89 & 0.27 & 0.97 & 0.62 \\
Cycling   & & 0.95 & 0.38 & 0.98 & 0.67 & & 0.95 & 0.15 & 0.97 & 0.66 \\
Walking   & & 0.93 & 0.86 & 0.95 & 0.88 & & 0.90 & 0.85 & 0.95 & 0.85 \\
\hline
\multicolumn{1}{l|}{\textbf{Nested logit}} & -3,285.18 &0.82 & & & &-1,157.64 & 0.81& \\
Car       & & 0.94 & 0.93 & 0.94 & 0.94 & & 0.93 & 0.92 & 0.94 & 0.93 \\
Bus       & & 0.90 & 0.64 & 0.94 & 0.79 & & 0.88 & 0.65 & 0.94 & 0.79 \\
Rail      & & 0.96 & 0.69 & 0.98 & 0.83 & & 0.94 & 0.64 & 0.97 & 0.80 \\
Taxi      & & 0.90 & 0.27 & 0.98 & 0.63 & & 0.89 & 0.27 & 0.97 & 0.62 \\
Cycling   & & 0.95 & 0.38 & 0.98 & 0.67 & & 0.95 & 0.15 & 0.97 & 0.66 \\
Walking   & & 0.93 & 0.86 & 0.95 & 0.88 & & 0.90 & 0.85 & 0.95 & 0.85 \\
\hline
\multicolumn{1}{l|}{\textbf{Asymmetric logit}} & -3,208.66 & 0.82& & & &-1,102.70 &0.81 & \\
Car       & & 0.94 & 0.94 & 0.94 & 0.94 & & 0.87 & 0.93 & 0.93 & 0.58 \\
Bus       & & 0.90 & 0.65 & 0.94 & 0.79 & & 0.88 & 0.65 & 0.94 & 0.79 \\
Rail      & & 0.96 & 0.69 & 0.98 & 0.83 & & 0.93 & 0.64 & 0.97 & 0.80 \\
Taxi      & & 0.90 & 0.28 & 0.98 & 0.63 & & 0.89 & 0.29 & 0.97 & 0.62 \\
Cycling   & & 0.95 & 0.39 & 0.98 & 0.67 & & 0.95 & 0.14 & 0.97 & 0.66 \\
Walking   & & 0.93 & 0.86 & 0.95 & 0.88 & & 0.90 & 0.85 & 0.95 & 0.85 \\
\hline
\multicolumn{1}{l|}{\textbf{Scobit}} & -3,236.97 &0.82 & & & &-1,100.54 &0.81 & \\
Car       & & 0.94 & 0.94 & 0.94 & 0.94 & &0.93 & 0.93 & 0.94 & 0.93 \\
Bus       & & 0.90 & 0.65 & 0.94 & 0.79 & &0.88 & 0.64 & 0.94 & 0.79 \\
Rail      & & 0.96 & 0.68 & 0.98 & 0.83 & &0.94 & 0.67 & 0.97 & 0.80 \\
Taxi      & & 0.95 & 0.27 & 0.98 & 0.62 & &0.95 & 0.28 & 0.97 & 0.62 \\
Cycling   & & 0.95 & 0.39 & 0.98 & 0.67 & &0.95 & 0.14 & 0.97 & 0.66 \\
Walking   & & 0.93 & 0.86 & 0.95 & 0.90 & &0.92 & 0.85 & 0.95 & 0.90 \\
\hline
\multicolumn{1}{l|}{\textbf{Uneven logit}} & -3,187.37&0.83 & & & &-1,124.37 &0.81 & \\
Car       & & 0.94 & 0.94 & 0.94 & 0.94 & & 0.93 & 0.93 & 0.94 & 0.93 \\
Bus       & & 0.91 & 0.65 & 0.94 & 0.80 & & 0.89 & 0.65 & 0.94 & 0.79 \\
Rail      & & 0.96 & 0.69 & 0.98 & 0.83 & & 0.94 & 0.64 & 0.97 & 0.80 \\
Taxi      & & 0.95 & 0.28 & 0.98 & 0.67 & & 0.95 & 0.29 & 0.97 & 0.66 \\
Cycling   & & 0.95 & 0.39 & 0.98 & 0.67 & & 0.95 & 0.14 & 0.97 & 0.66 \\
Walking   & & 0.93 & 0.86 & 0.95 & 0.88 & & 0.90 & 0.85 & 0.95 & 0.85 \\
\hline
\multicolumn{1}{l|}{\textbf{Random forest}} & -3,150.73 & 0.86 & & & & -1,125.36 &0.86 & \\
Car       & & 0.89 & 0.86 & 0.92 & 0.89 & & 0.89 & 0.86 & 0.92 & 0.89 \\
Bus       & & 0.88 & 0.61 & 0.92 & 0.77 & & 0.86 & 0.60 & 0.91 & 0.75 \\
Rail      & & 0.94 & 0.50 & 0.96 & 0.73 & & 0.95 & 0.50 & 0.96 & 0.73 \\
Taxi      & & 0.94 & 0.11 & 0.97 & 0.54 & & 0.94 & 0.10 & 0.97 & 0.54 \\
Cycling   & & 0.94 & 0.25 & 0.96 & 0.60 & & 0.95 & 0.17 & 0.96 & 0.56 \\
Walking   & & 0.91 & 0.81 & 0.94 & 0.87 & & 0.89 & 0.77 & 0.94 & 0.86 \\
\hline
\multicolumn{1}{l|}{\textbf{Artificial neural network}} & -3,171.66 & 0.88 & & & &-1,138.01 &0.86 & \\
Car       & & 0.91 & 0.88 & 0.93 & 0.90 & & 0.90 & 0.87 & 0.92 & 0.89 \\
Bus       & & 0.90 & 0.65 & 0.94 & 0.79 & & 0.87 & 0.62 & 0.92 & 0.77 \\
Rail      & & 0.96 & 0.65 & 0.97 & 0.81 & & 0.96 & 0.58 & 0.97 & 0.77 \\
Taxi      & & 0.94 & 0.15 & 0.97 & 0.56 & & 0.94 & 0.17 & 0.97 & 0.57 \\
Cycling   & & 0.95 & 0.38 & 0.97 & 0.67 & & 0.95 & 0.16 & 0.97 & 0.56 \\
Walking   & & 0.91 & 0.83 & 0.94 & 0.88 & & 0.90 & 0.80 & 0.94 & 0.87 \\
\hline
\multicolumn{1}{l|}{\textbf{XGBoost}} & -3,136.17 & 0.88 & & & & -1,149.08 & 0.86& \\
Car       & & 0.90 & 0.87 & 0.93 & 0.90 & & 0.90 & 0.87 & 0.92 & 0.89 \\
Bus       & & 0.89 & 0.65 & 0.93 & 0.79 & & 0.87 & 0.62 & 0.92 & 0.77 \\
Rail      & & 0.95 & 0.62 & 0.97 & 0.79 & & 0.96 & 0.58 & 0.97 & 0.77 \\
Taxi      & & 0.95 & 0.18 & 0.97 & 0.58 & & 0.94 & 0.17 & 0.97 & 0.57 \\
Cycling   & & 0.95 & 0.38 & 0.97 & 0.67 & & 0.95 & 0.16 & 0.97 & 0.56 \\
Walking   & & 0.91 & 0.83 & 0.94 & 0.89 & & 0.90 & 0.80 & 0.94 & 0.87 \\

\end{tabular}
\label{tab:metrics_decisions}
\end{table}

\newgeometry{margin=1.5cm}
\begin{landscape}
\begin{figure}[p]
\centering
\cmpanel{decsubfig}{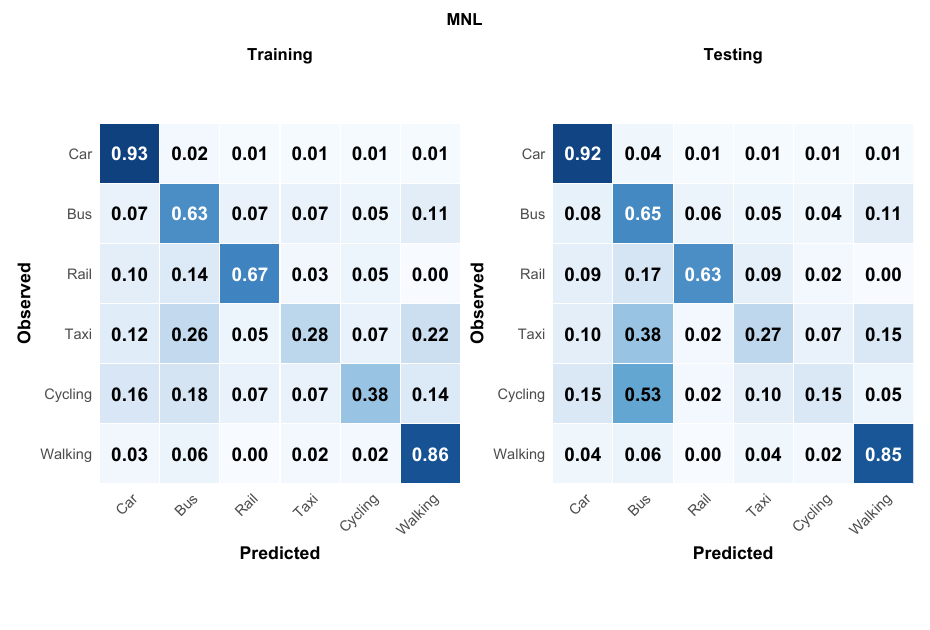}{MNL}{Decisions}{fig:dec_4A}\hfill
\cmpanel{decsubfig}{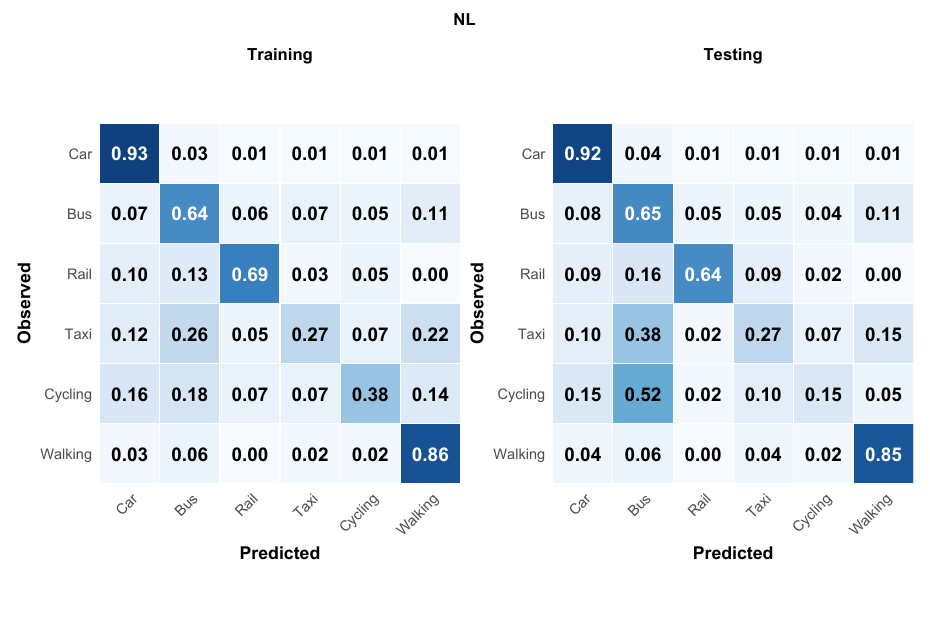}{NL}{Decisions}{fig:dec_4B}\hfill
\cmpanel{decsubfig}{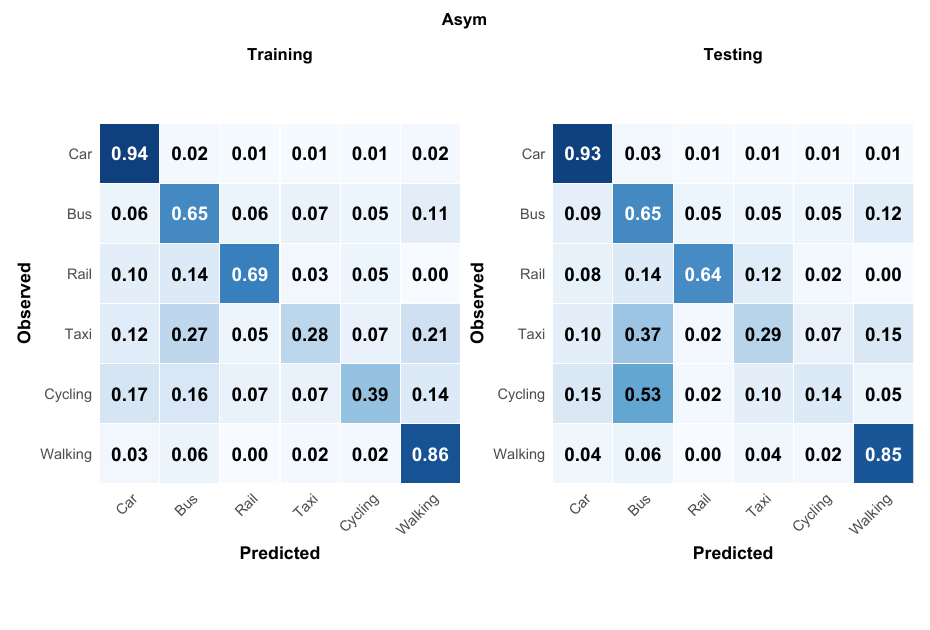}{Asymmetric}{Decisions}{fig:dec_4C}\hfill
\cmpanel{decsubfig}{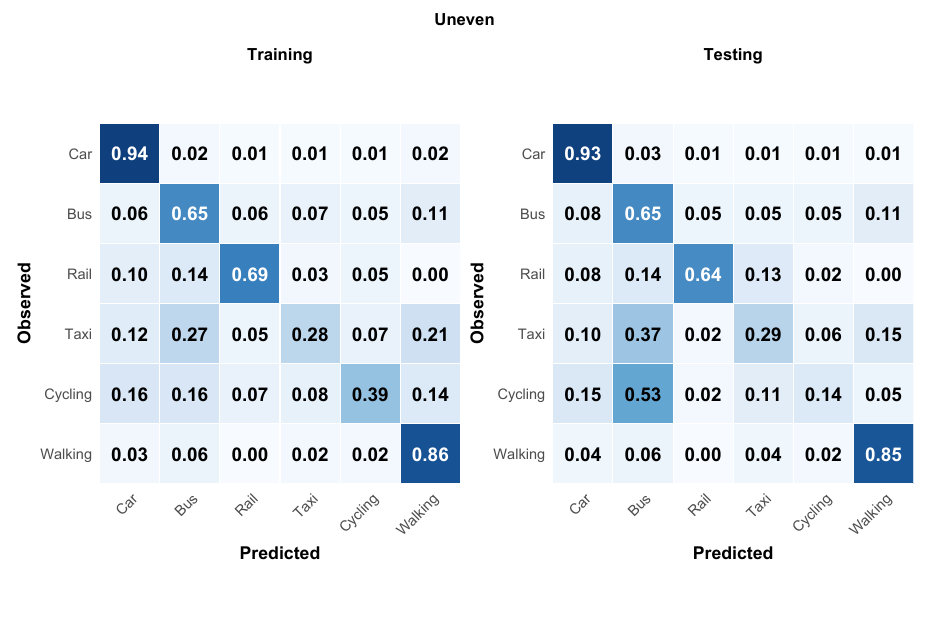}{Uneven}{Decisions}{fig:dec_4D}

\vspace{1.5em}

\cmpanel{decsubfig}{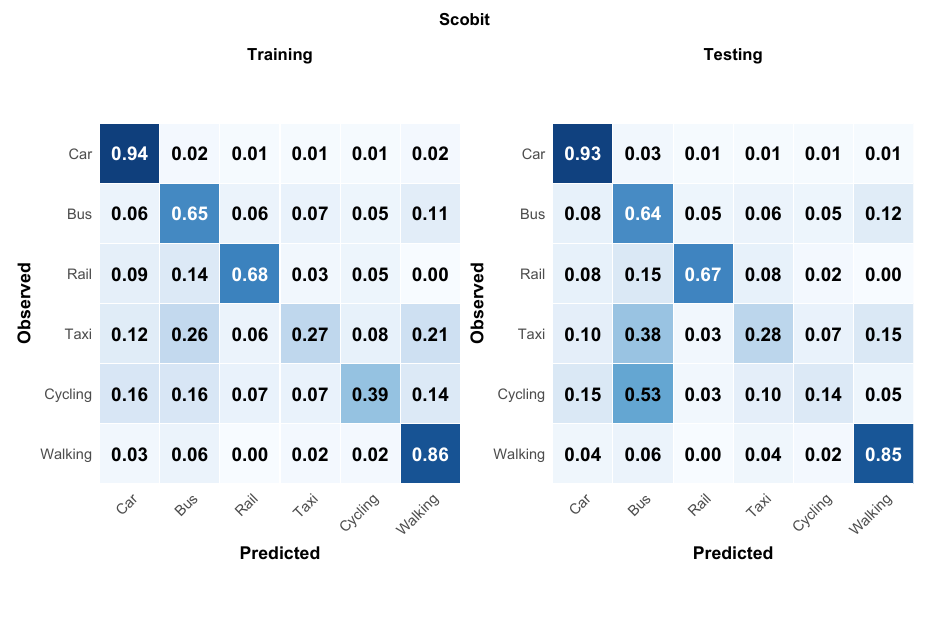}{Scobit}{Decisions}{fig:dec_4E}\hfill
\cmpanel{decsubfig}{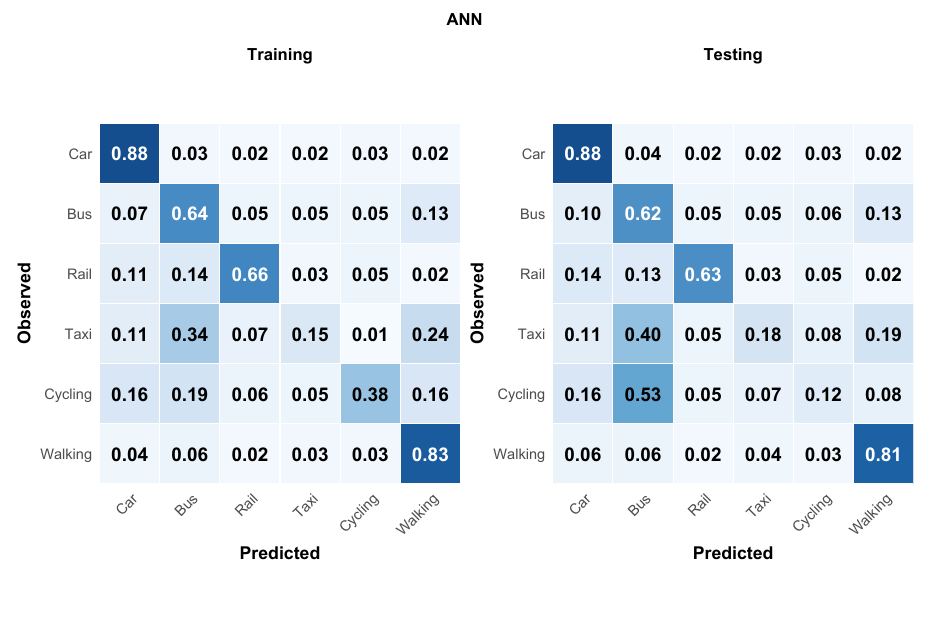}{ANN}{Decisions}{fig:dec_4F}\hfill
\cmpanel{decsubfig}{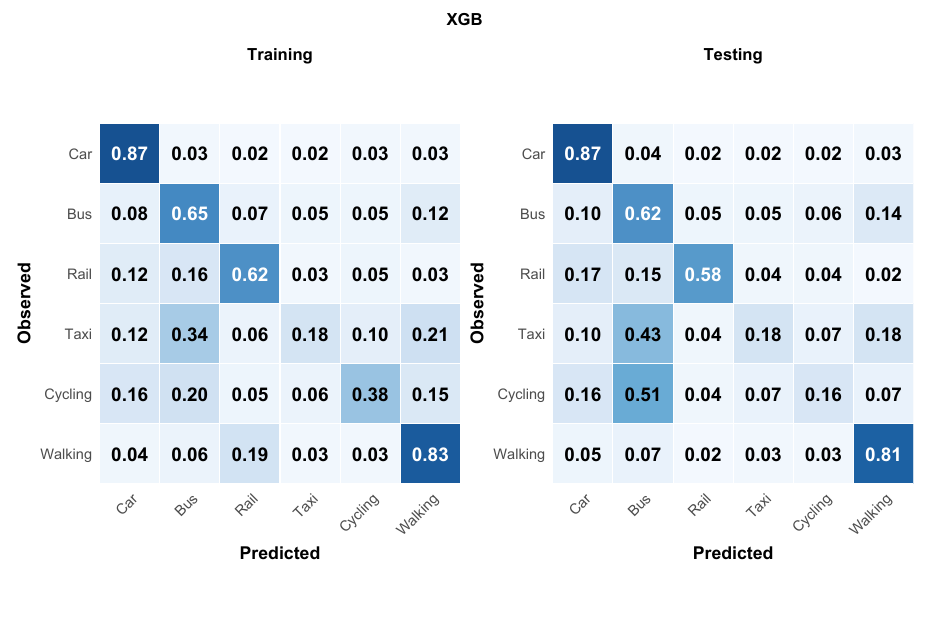}{XGB}{Decisions}{fig:dec_4G}\hfill
\cmpanel{decsubfig}{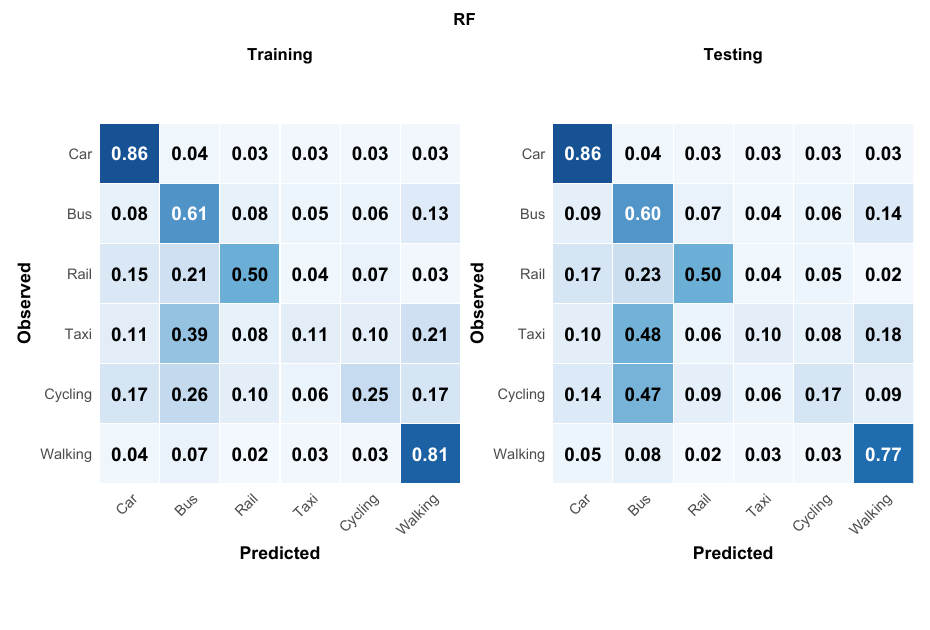}{RF}{Decisions}{fig:dec_4H}
\end{figure}
\end{landscape}
\restoregeometry

\section{Conclusions and next steps}
\label{sec:conclusion}

In this paper, we provide a new benchmark for evaluating choice models by analysing their predictive performance through tools commonly used in ML. We extend the traditional confusion matrix into a probabilistic form, enabling the evaluation of probability distributions rather than deterministic outputs, and show how this matrix can be used to compute probabilistic ML metrics: accuracy, sensitivity, specificity, and balanced accuracy. These metrics complement classical CM evaluation criteria, such as log-likelihood, and provide deeper insights into alternative-level performance of models, which cannot be obtained from aggregate metrics alone. To the best of our knowledge, this is the first study of these characteristics, also bridging methodological practices across CM and ML.  

We conclude that, although these metrics should not be used as a sole criterion for model selection, they provide valuable complementary diagnostic information, as they offer a more nuanced view of model performance by: (1) disaggregating results by alternative, revealing which alternatives are poorly predicted and hence where specification improvements should be targeted; (2) penalising performance for incorrect predictions, as FP and FN also contribute to model performance; (3) revealing misclassification patterns, by showing how probabilities are distributed across alternatives when the observed alternative is not well predicted; and (4) helping identify potential overfitting, as discrepancies between training and testing confusion matrices highlight where models fit the training data too closely and fail to generalise. 

Given the large number of studies claiming that ML outperforms CM because it achieves a higher accuracy, our results suggest that accuracy alone is an insufficient and often misleading criterion, especially in imbalanced datasets. High accuracy can be achieved even when a model performs poorly on minority alternatives, as this metric can be dominated by TN, rather than the correct identification of the chosen alternative. We also suggest that choice modellers using ML classifiers prioritise probabilistic outputs over deterministic predictions. Probabilistic evaluation accounts for the full distribution of the predicted probabilities, not just the highest probability, and thus preserves information about the non-chosen alternatives. Conversely, deterministic outputs often yield zero predictions for rarely chosen alternatives, thereby inflating accuracy. 

This work is not exempt from limitations. Our probabilistic ML metrics sometimes produce results that are similar across modes, particularly when the underlying choice probabilities are well captured by standard specifications. In such cases, more complex structures do not necessarily translate into substantial improvements in the confusion matrices. Therefore, we encourage other researchers to apply these methods to other datasets to effectively understand if there are greater differences between models for other scenarios. 

An important future step is to use these metrics to guide model specification. Off-diagonal patterns and sensitivities in the confusion matrices provide direct insights into where a model is underperforming. For example, as highlighted in \autoref{subsec:case_decisions}, the Decisions dataset shows that Bus receives high probabilities for Taxi and Cycling observations. An immediate next step would be to refine Bus' utility function to reduce these mispredictions. Furthermore, a promising line of research is automating model specification by using the probabilistic confusion matrix entries as optimisation criteria (cf. \cite{ortelli2021, nova2025}). This would be a multi-objective optimisation, in which the goals are to minimise incorrect classifications and maximise the diagonal elements. Inputs for such a process could include the set of alternatives, their attributes, potential attribute transformations (e.g., logarithmic or Box-Cox), respondents' socioeconomic characteristics, and model structures (e.g., NL or Uneven).

\section*{Acknowledgments}
Lorenzo Mu\~{n}oz, Stephane Hess, Thomas Hancock, and Georges Sfeir acknowledge the financial support by the European Research Council through the advanced grant 101020940-SYNERGY.

\bibliography{references}

\appendix

\section{Modelling background} 
\label{appendix:modelling_background}

\subsection{Choice models} 
\label{subsec_ap:choicemodels}

In this section, we describe the key differences between models for the probability computation. We first estimate a multinomial logit (MNL) incorporating alternative-specific constants ($\alpha_i$), alternatives' attributes and respondents' covariates ($x_{int}$), following \autoref{eq:utility}:

\begin{equation}
U_{in}=V_{in}+\epsilon_{in}=\alpha_{i}+\beta_{i}x_{in}+\epsilon_{in}
\label{eq:utility}
\end{equation}

We also turn to more complex specifications. The Nested Logit model (NL; \cite{nested, nesteddaly}) extends the MNL by allowing different alternatives to  be grouped into hierarchical structures, or `nests'. Within each nest, alternatives share a common unobserved component, relaxing the independence of irrelevant alternatives assumption. For the NL, the probability of individual $n$ choosing alternative $i$ that belongs to nest $m$ with a scale parameter $\lambda_m$ is given by \autoref{eq:nl}:

\begin{equation}
P_{in} = P_{mn}P(i \mid m)_{n},
\label{eq:nl}
\end{equation}

where 

\begin{equation}
P(i \mid m)_{n} =
\frac{e^{\frac{V_{in}}{\lambda_m}}}
{\sum\limits_{j \in m} e^{\frac{V_{jn}} {\lambda_m}}}
\end{equation}

and

\begin{equation}
P_{mn} =
\frac{e^{\frac{\lambda_m}{\lambda_r}}}
{\sum\limits_{j \in m} e^{\frac{1} {\lambda_m}}}
\end{equation}
\par

where $\lambda_r$ is the nest parameter for an upper level nest, i.e., $0\leq\lambda_m\leq\lambda_r\leq1$. In a two-level NL, $\lambda_r=1$.

We also estimate models that assume changes in the symmetry of the probability distribution: Scobit, Uneven and Asymmetric models, as introduced by \cite{brathwaite2018}. These models account for skewed response behaviour by considering the aforementioned scale parameters $\delta_i$ in the probability function, differing in their specifications. By estimating these models, we also address an important methodological gap at the intersection of CM and ML: most comparisons are usually applied to an MNL, the simplest logit model \citep{salas2022}.

The probability of choosing alternative $i$ in a multinomial Scobit model is given by \autoref{eq:scobit}, where each alternative has an associated estimated scalar $\gamma_i$.

\begin{equation}
P_{in}=\frac{exp(\delta_i-ln[(1+e^{-V_{in}})^{\gamma_i}]-1)}
{\sum\limits_{j=1}^{I} {exp(\delta_j-ln[(1+e^{-V_{jn}})^{\gamma_j}]-1)}}
\label{eq:scobit}
\end{equation}

The probability of choosing alternative $i$ in an Uneven logit model is given by \autoref{eq:uneven}, where each alternative has an associated estimated shape parameter $\tau_i$.

\begin{equation}
P_{in}=\frac{exp[\delta_i+V_{in}+ln(1+e^{-V_{in}})^{\tau_i}-ln(1+e^{-{\tau_i}V_{in}})]}
{\sum\limits_{j=1}^{I} {exp[\delta_j+V_{jn}+ln(1+e^{-V_{jn}})^{\tau_j}-ln(1+e^{-{\tau_j}V_{jn}})]}}
\label{eq:uneven}
\end{equation}

The probability of choosing an alternative $i$ in an Asymmetric logit model is given by \autoref{eq:asym}, where the shape parameters have the restrictions: $0\leq \kappa_i \leq 1$ and ${\sum\limits_{I} {\kappa_i=1}}$.

\begin{equation}
P_{in}=\frac{exp[\delta_i+S_{in}]}
{\sum\limits_{j=1}^{I} {exp[\delta_j+S_{jn}]}}
\label{eq:asym}
\end{equation}

\[
S_{in} =
\begin{cases} 
  ln(\kappa_i)-V_{in}ln(\kappa_i), & \text{if } V_{in} \geq 0 \\
  ln(\kappa_i)-V_{in}ln(\frac{1-\kappa_i}{I-1}), & \text{if } V_{in} < 0
\end{cases}
\]

All the aforementioned models were estimated with the R package Apollo \citep{hess2019apollo}.

\subsection{Machine learning algorithms} 
\label{subsec_ap:ml}

Methods from the ML field offer a promising alternative for modelling choices \citep{hagenauer2017}. Unlike CM---which impose a predetermined, often linear utility specification---ML algorithms allow for highly flexible functional forms, potentially reducing misspecification risks and improving the model's ability to fit empirical data \citep{xie2003}. Another advantage is that ML models do not require full information on all non-chosen alternatives to generate predictions, whereas CM rely explicitly on the attributes of both chosen and competing alternatives \citep{zhao2020}. However, this is not ideal, as it is inconsistent with the underlying econometric theory in CM. 

XGB is a gradient-boosting method that builds an ensemble of decision trees sequentially (cf. \cite{chen2016xgboost}). Each new tree is trained to correct the residual errors (i.e., the gradients of the loss function) from the previous trees, effectively `boosting' the model’s predictive performance \citep{salas2022}. Hard to predict observations receive greater weight in later iterations. \autoref{eq:xgb1} shows the probability $P_{int}$ for an alternative $i$ and individual $n$, where $f_k$ is the $k$-th decision tree from the set $K$, with $\mathcal{F}$ being the space of all regression trees. 

\begin{equation}
P_{in} = \sum_{k=1}^{K} f_k(x_{int}), \quad f_k \in \mathcal{F}
\label{eq:xgb1}
\end{equation}

Training minimises a regularised objective that combines a differentiable loss with penalties on tree complexity to prevent overfitting. In XGB, optimisation can be performed using different objectives, including multi-alternative cross-entropy (which is equivalent to log-likelihood), logistic loss, and squared error. These loss functions are probabilistic in nature and operate on the full vector of predicted probabilities. The key difference is that, while CM explicitly maximises the log-likelihood of the chosen alternative, ML loss functions typically penalise discrepancies across the entire predicted probability distribution. This allows XGB to flexibly capture non-linear relationships while remaining computationally efficient. 

ANNs model choice behaviour by combining interconnected artificial neurons that capture non-linear relationships between variables (cf. \cite{rojas2013neural}). We employ a Feed-Forward Neural Network (FFNN), also known as a Multi-Layer Perceptron (MLP), which consists of an input layer, one or more hidden layers, and an output layer with one node per alternative. The output layer produces alternative scores (logits)---analogous to CM utilities---and probabilities are obtained using a Softmax function---analogous to the MNL probability function. The hidden layer transformation for an observation $O$ is presented in \autoref{eq:ann}, where the feature vector $x_{int}$ is transformed using a weight matrix $W$ and a bias term $b$. 

\begin{equation}
h_{o} = g(W x_{int} + b)
\label{eq:ann}
\end{equation}

Here, $g(\cdot)$ denotes a generic activation function. Different activation functions can be used---such as ReLU, tanh, or sigmoid---and the choice is typically determined through hyperparameter tuning. This flexibility allows the network to capture a wide range of non-linear patterns in the data. 

RF is a tree-based ensemble that trains multiple classification trees in parallel using bootstrap samples (cf. \cite{breiman2001}). At each split, a random subset of features is considered, reducing correlation between trees and improving generalisation. RF offers several advantages: it is relatively interpretable, as predictions can be traced through individual decision trees composed of transparent sequences of decision rules, allowing the modeller to inspect which variables drive specific classification outcomes. It also handles qualitative predictors without dummy encoding, manages imbalanced datasets effectively, and is computationally efficient \citep{cottreau2025}. Each tree $r \in R$ outputs the predictions of choice probabilities, then, the RF predicted probability is given in \autoref{eq:rf}:

\begin{equation}
P_{int} = \frac{1}{R} \sum_{r=1}^{R} R_r(x_{int}).
\label{eq:rf}
\end{equation}

An important difference between most ML outputs and econometric utilities is that, frequently in ML, the modeller does not control which variables correspond to which alternative-specific utility. All available features are fed into the model simultaneously, meaning that variables associated with one alternative may influence the predicted score for another. Moreover, often the availability of alternatives is not used as a $constraint$ as in CM, meaning that unavailable alternatives can have non-zero probabilities, which lacks behavioural interpretation. Instead, availabilities are treated as $features$. Specifically, this could have implications on off-diagonal elements. For instance, we note that Car FP are higher for ML algorithms than for CM structures (see \autoref{subsec:case_decisions}).

\section{Overfitting check} 
\label{appendix:overfitting}

\begin{figure}[h!]
  \centering
  {%
  \captionsetup{font=footnotesize}%
  \begin{minipage}{0.48\textwidth}\centering
    \includegraphics[width=\linewidth]{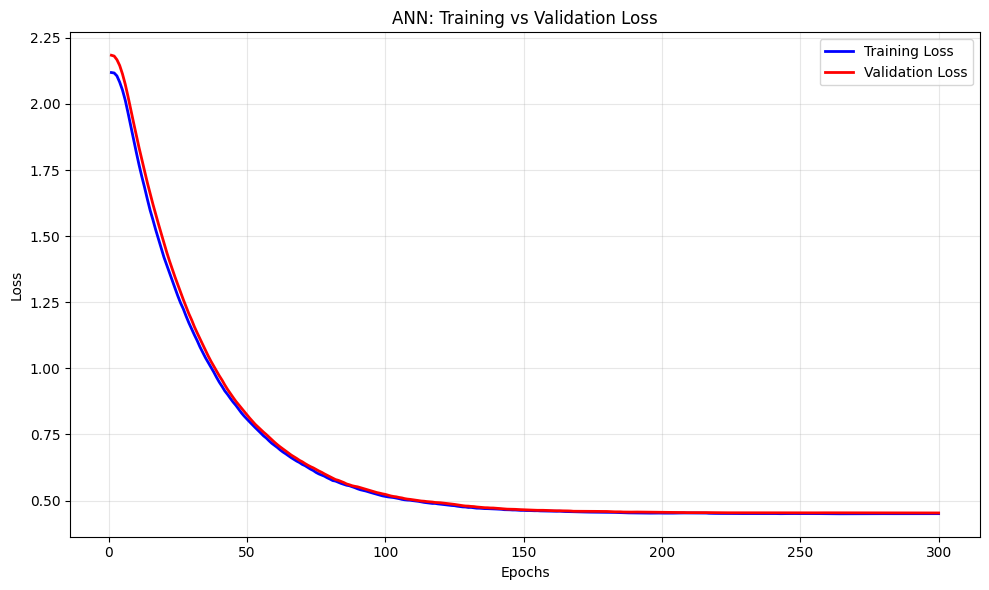}
    \caption{overfitting check, Covid ANN.}
    \label{fig:overfit_ann_covid}
  \end{minipage}\hfill
  \begin{minipage}{0.48\textwidth}\centering
    \includegraphics[width=\linewidth]{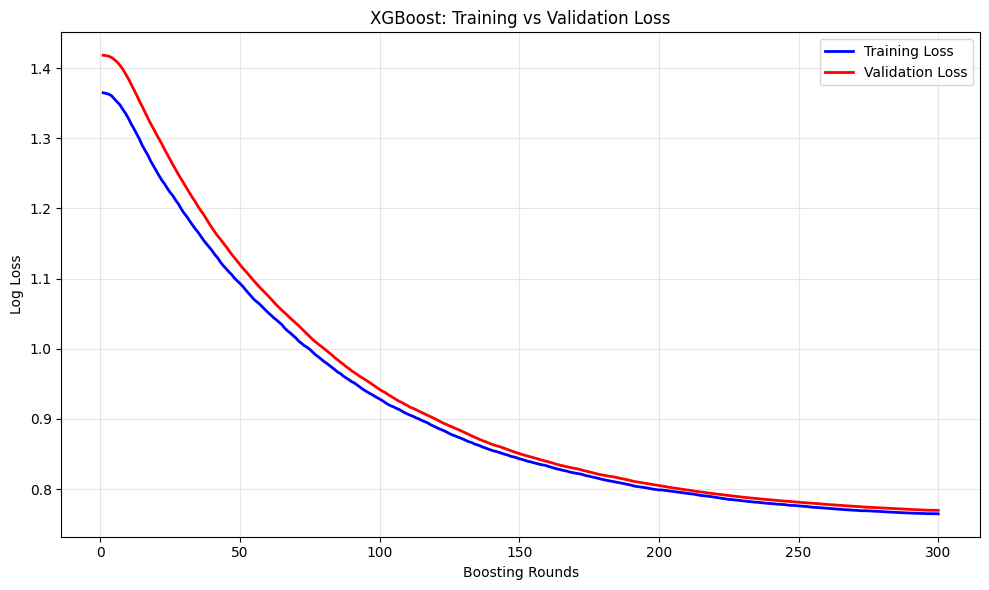}
    \caption{overfitting check, Covid XGB.}
  \end{minipage}%
  }
\end{figure}

In this section, we present the overfitting diagnostics for the evaluated models. As previously described, a 20\% validation set was retained from the training data, and hyperparameter tuning was systematically conducted across all machine learning models. The search spaces and final optimal configurations for each dataset are reported in \autoref{tab:hyperparameters}. Thus, we present the plots of the training and validation losses for ANN and XGB across all datasets, from \autoref{fig:overfit_ann_covid} to \autoref{fig:overfit_decisions_xgb}. In all instances, the training and validation loss curves converge and follow closely aligned paths, indicating that the models do not exhibit signs of overfitting. For RF, which does not train iteratively over epochs, overfitting was controlled through structural regularisation hyperparameters (such as maximum depth and minimum samples per leaf) alongside bootstrap aggregation, with generalisation monitored on the validation set.

\begin{table}[htbp]
\centering
\small
\caption{Hyperparameter search spaces and final configurations for all datasets.}
\begin{tabular}{llcccc}
\toprule
\textbf{Model} & \textbf{Hyperparameter} & \textbf{Search Space} & \textbf{Covid} & \textbf{Swissmetro} & \textbf{Decisions} \\
\midrule
\multicolumn{6}{l}{\textbf{ANN}} \\
& Hidden Layers & $[1, 5]$ & 4 & 4 & 1 \\
& Layer Units & $[32, 512]$ & [104, 104, 128, 128] & [80, 48, 80] & 192 \\
& Dropout Rate & $[0.0, 0.5]$ & 0.0 & [0.4, 0.0, 0.2] & 0.4 \\
& L1 Penalty ($\alpha$) & $[10^{-6}, 0.1]$ & 0.046 / 0.040 / 0.009 & 0.0 & 0.0 \\
& L2 Penalty ($\lambda$) & $[10^{-5}, 0.1]$ & 0.016 / 0.044 / 0.005 & 0.0 & 0.0 \\
& Batch Normalisation & \{\text{True, False}\} & False & False & True \\
& Learning Rate & $[10^{-4}, 10^{-2}]$ & 0.0079 (AdamW) & 0.0010 (Adam) & 0.0010 (Adam) \\
& Optimizer Parameters & -- & $\beta_1=0.89, \beta_2=0.99$ & Default & Default \\
\midrule
\multicolumn{6}{l}{\textbf{XGB}} \\
& Number of Estimators & $[100, 1050]$ & 200 & 1,050 & 549 \\
& Max Depth & $[3, 30]$ & 30 & 7 & 7 \\
& Learning Rate & $[0.005, 0.2]$ & 0.020 & 0.005 & 0.033 \\
& Subsample Ratio & $[0.5, 1.0]$ & 0.500 & 0.700 & 0.830 \\
& Colsample by Tree & $[0.5, 1.0]$ & 0.500 & 0.604 & 0.902 \\
& Min Child Weight & $[1, 20]$ & 3 & 20 & 1 \\
& Gamma ($\gamma$) & $[0.0, 10.0]$ & 0.500 & 10.000 & 0.207 \\
& L1 Regularisation ($\alpha$) & $[0.0, 5.0]$ & 0.100 & 5.000 & 0.037 \\
& L2 Regularisation ($\lambda$) & $[0.5, 10.0]$ & 1.000 & 10.000 & 0.565 \\
\midrule
\multicolumn{6}{l}{\textbf{RF}} \\
& Number of Trees & $[100, 1000]$ & 600 & 500 & 488 \\
& Max Depth & $[5,10,30, \text{None}]$ & 30 & 30 & 10 \\
& Min Samples Split & $[2, 20]$ & 10 & 10 & 2 \\
& Min Samples Leaf & $[1, 10]$ & 2 & 4 & 4 \\
& Max Feature Ratio & $[0.1, 1.0]$ & 0.5 & 0.4 & 0.7 \\
& Min Weight Fraction Leaf & $[0.0, 0.01]$ & 0.0012 & 0.0000 & 0.0012 \\
& Bootstrap Sampling & \{\text{True, False}\} & True & True & True \\
\bottomrule
\end{tabular}
\label{tab:hyperparameters}
\end{table}

\begin{figure}[h!]
  \centering
  {%
  \captionsetup{font=footnotesize}%
  \begin{minipage}{0.48\textwidth}\centering
    \includegraphics[width=\linewidth]{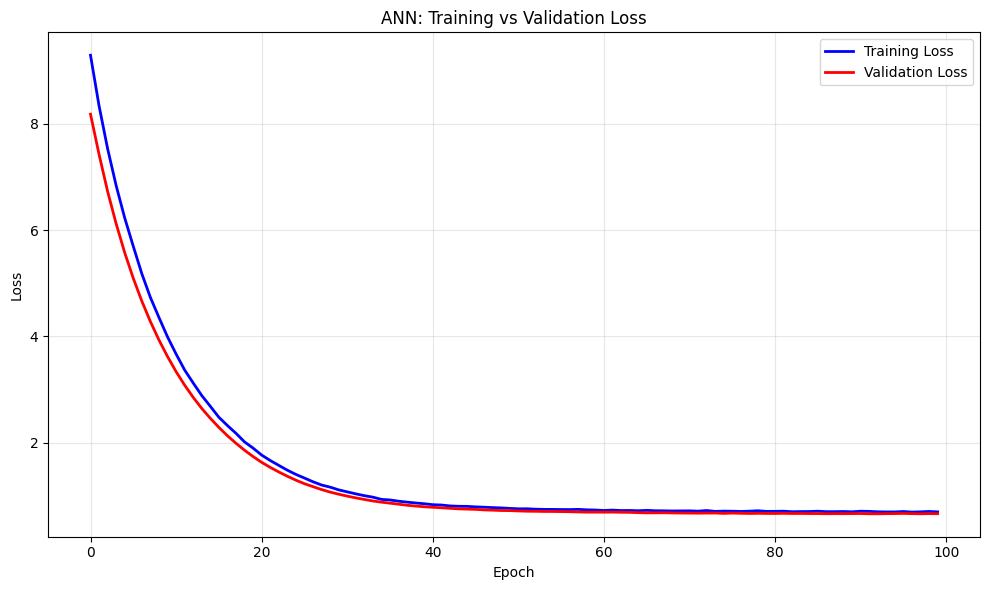}
    \caption{overfitting check, SwissMetro ANN.}
  \end{minipage}\hfill
  \begin{minipage}{0.48\textwidth}\centering
    \includegraphics[width=\linewidth]{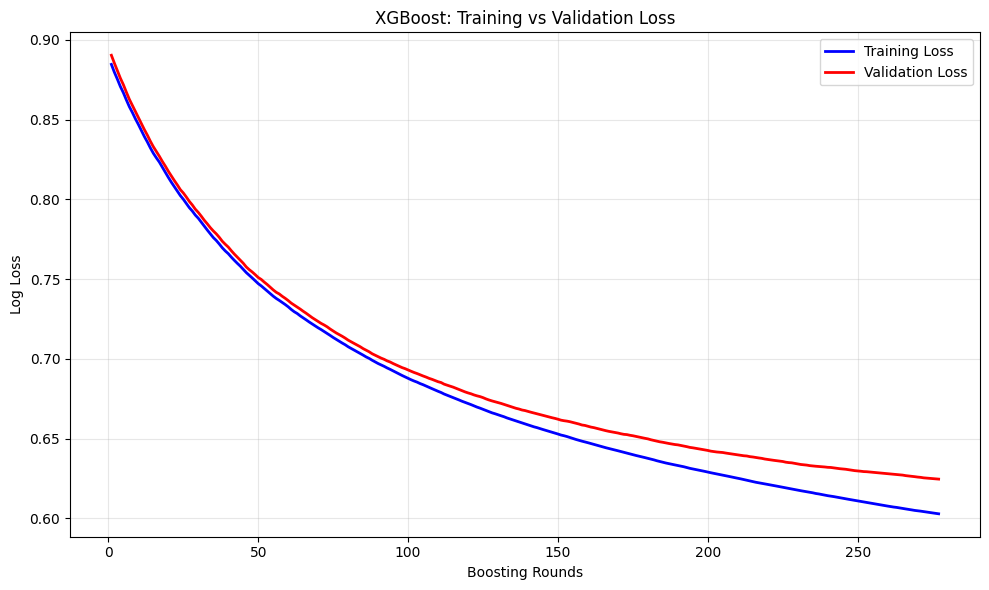}
    \caption{overfitting check, SwissMetro XGB.}
  \end{minipage}%
  }
\end{figure}

\begin{figure}[h!]
  \centering
  {%
  \captionsetup{font=footnotesize}%
  \begin{minipage}{0.48\textwidth}\centering
    \includegraphics[width=\linewidth]{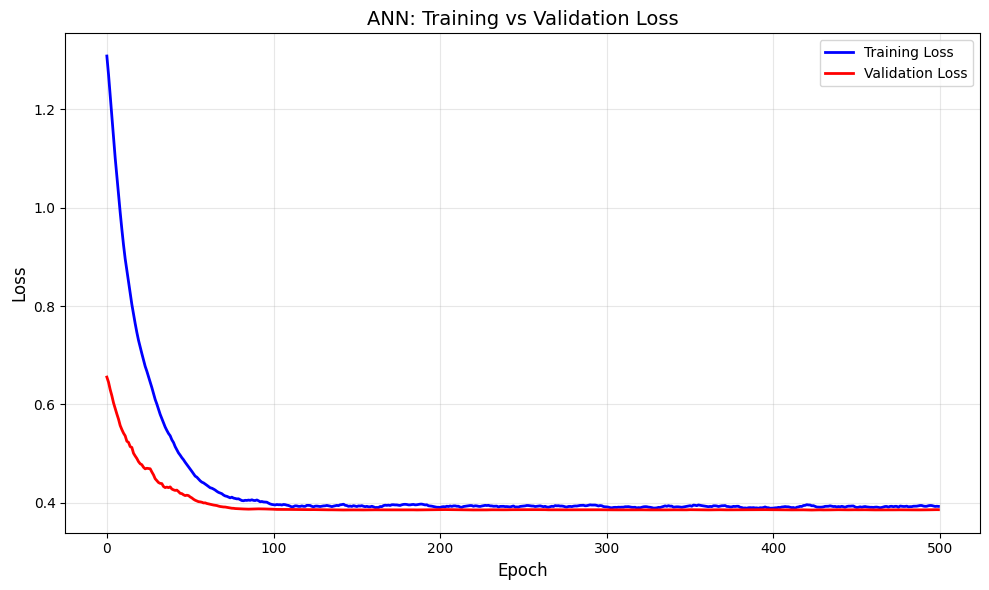}
    \caption{overfitting check, Decisions ANN.}
  \end{minipage}\hfill
  \begin{minipage}{0.48\textwidth}\centering
    \includegraphics[width=\linewidth]{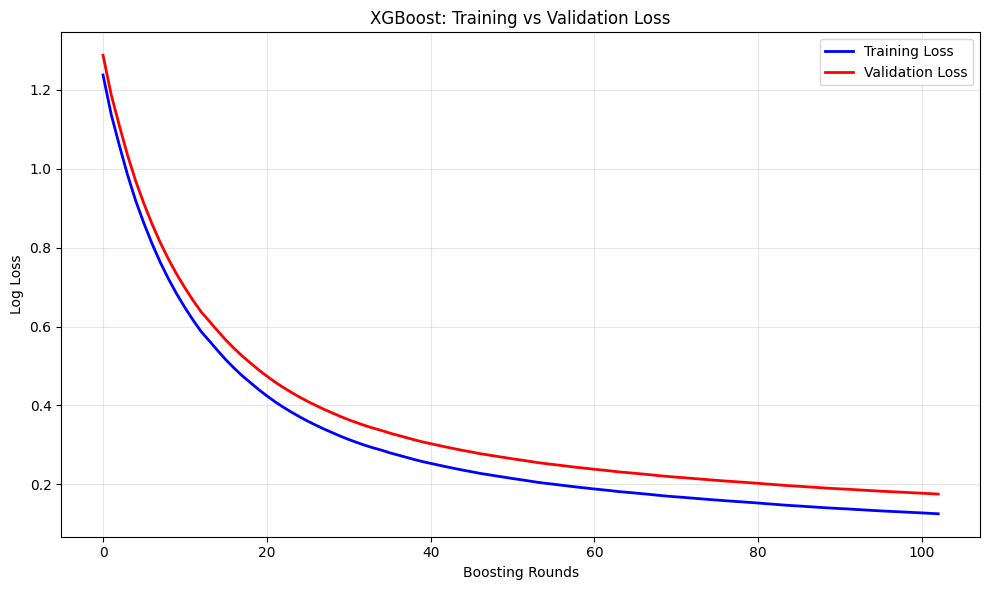}
    \caption{overfitting check, Decisions XGB.}
    \label{fig:overfit_decisions_xgb}
  \end{minipage}%
  }
\end{figure}

\end{document}